\documentclass[aps,prb,showpacs,amsmath,superscriptaddress,twocolumn,floatfix,nobalancelastpage]{revtex4-1}

\usepackage[nottoc,numbib]{tocbibind}

\usepackage{mathrsfs} 

\usepackage[colorlinks=true, citecolor=black, linkcolor=black, urlcolor=black]{hyperref}

\usepackage[all]{hypcap}
\usepackage{amsmath}
\usepackage{enumerate,bm}
\usepackage{graphicx}
\usepackage{grffile} % to handle dots in graphics filnames
\usepackage{color}
\usepackage{esint}
\usepackage{textpos}
\usepackage[usenames,dvipsnames]{xcolor}
\usepackage{subfigure}
\usepackage{flushend}
\usepackage{tabularx}
\usepackage{amssymb}
\usepackage{float}
\usepackage{subfigure}
\usepackage{ulem} %striketrhough

\usepackage{lineno} % line numbers in pdf
\usepackage{setspace}
\usepackage{verbatim}

\begin{document}

\author{Umberto~De~Giovannini}
\email{umberto.degiovannini@gmail.com}
\affiliation{Max Planck Institute for the Structure and Dynamics of Matter and Center for Free Electron Laser Science, 22761 Hamburg, Germany}
%\affiliation{Dipartimento di Fisica e Chimica, Universit\'a degli Studi di Palermo, Via Archirafi 36, I-90123, Palermo, Italy}

\author{Hannes~H\"ubener}
\email{hannes.huebener@gmail.com}
\affiliation{Max Planck Institute for the Structure and Dynamics of Matter and Center for Free Electron Laser Science, 22761 Hamburg, Germany}

\title{Floquet analysis of excitations in materials}

\begin{abstract}
    Controlled excitation of materials can transiently induce changed or novel properties with many fundamental and technological implications. Especially, the concept of Floquet engineering, manipulation of the electronic structure via dressing with external lasers, has attracted some recent interest. Here we review the progress made in defining Floquet materials properties and give a special focus on their signatures in experimental observables as well as considering recent experiments realizing Floquet phases in solid state materials. We discuss how a wide range of experiments with non-equilibrium electronic structure can be viewed by employing Floquet theory as an analysis tool providing a different view of excitations in solids.   
\end{abstract}

\maketitle

\section{Introduction}
Controlling and manipulating materials properties by driving them out of equilibrium is fast emerging as an exciting field of research\cite{Hsieh:2017ix}. Prominent examples include signatures of light-induced\cite{Mitrano:2016fr} or light enhanced\cite{Mankowsky:2014em} superconductivity, ultrafast switching of hidden electronic phases\cite{Stojchevska:2014it} and phonon induced magnetization\cite{Nova:2017ja}. In particular the concept of viewing the non-equilibrium electronic structures as dressed by external fields and thus allowing the control of so called synthetic gauge fields\cite{Aidelsburger:2018jf} has attracted much attention because it shows a route towards controlling topological and other properties via a process called Floquet engineering\cite{Oka:2018df}. Based on the theory of differential equations with oscillating coefficients, Floquet phases are thought to occur when a quantum mechanical system is subjected to a periodically oscillating potential, such as for instance, a laser.

Many intriguing non-equilibrium properties have been induced in a controlled way by applying oscillating perturbations to artificial lattices in so called quantum simulation settings\cite{Eckardt:2017hc} including Floquet topology\cite{Rechtsman:2013fe,Jotzu:2014kz}. While in real materials realizations of Floquet engineered phases are still far and few between, we concentrate here only on Floquet phases for condensed matter systems. Observation of non-equilibrium phases is a non-trivial endeavour and a field in itself\cite{Wang:2018fy}, however here we discuss a few examples of observation of Floquet states or at least of observables resulting from Floquet phases. It is important to note that realistic Floquet phases are intricately linked to many-body\cite{Eisert:2015ka} and thermodynamical aspects\cite{Moessner:2017jb} which can support or obfuscate a pure Floquet picture. 

Besides reviewing proposals and successful observations of Floquet phases we also explore the application of Floquet theory to solid state systems in a broader sense. The basic requirement for Floquet theory to be applicable is a periodically oscillating potential. Most proposals of Floquet phases to date consider this to be an externally applied field or other perturbation. Instead, an oscillating potential is also provided by internal modes of a material. Usually such modes are associated with quasi-particles that themselves have a dispersion relation and thus do not necessarily provide a single frequency. Therefore, such modes need to be selectively excited to be the dominating term, thus creating an non-equilibrium dressed electronic structure that is intrinsic to the material and viewing it with Floquet theory opens a new perspective on materials properties. 

This idea, of employing Floquet theory to perform analysis of materials excitations, equally applies to externally driven phases. We will show how comparing non-equilibrium observables of the Floquet electronic structure can lead to valuable insight into experiments. Here Floquet theory is not viewed as a way of constructing analytically a non-equilibrium state, but as a theoretical tool to understand certain states of matter under non-equilibrium conditions. In this sense Floquet states are considered an idealization of a more complex physical reality. How much of this idealization can be found in experiments remains to be seen and we propose to employ Floquet theory as a tool of analysis of non-equilibrium states.

In the following we first review the basic theoretical concepts of Floquet theory when applied to quantum mechanics, specifically the time-dependent Schr\"odinger equation. We then move to discuss a variety of observables of Floquet systems and their formulation within Floquet theory, forming the basic tools for Floquet analysis. The interpretation framework that Floquet analysis provides is then illustrated by explicitly solving a driven two level system and discussing its time-dependent properties in terms of Floquet theory. After this first part that lays out the theoretical background, we discuss in the second part various theoretical proposals for Floquet engineered materials as well as experimental realizations. We aim to give a broad overview over various flavours of Floquet material and provide a rough categorization. Finally we discuss how Floquet analysis can be used to look at internal modes and properties of materials, through dressing mechanism beyond electromagnetic fields.     

\section{Theoretical framework}
\subsection{General Theory}

\textbf{Floquet states}. Floquet theory is an area of mathematics dating back to the work of G. Floquet in 1883 that deals with first order differential equations with periodic coefficients\cite{Floquet:1883eo}. Its most important finding is often called the Floquet-Lyapunov theorem and states that while solutions to such differential equations generally do not have the same periodicity as the coefficients, they can nevertheless be expressed as the product of a function with that periodicity and a constant phase. The specialization of this theorem to the static Schr\"odinger equation with periodic lattice potentials is well known in solid states physics as the Bloch-theorem or Floquet-Bloch theorem. However, here we are interested in the time-dependent Schr\"odinger equation as a differential equation in time
\begin{equation}
    i\frac{\partial }{\partial t}\psi(t) = H(t)\psi(t)
\end{equation}
with the condition that $H(t+T)=H(t)$ and we call $\Omega = 2\pi/T$ the Floquet frequency. It then follows from the Floquet theorem that we can write any solution of the Schr\"odinger equation as a linear combination\cite{Shirley:1965cy}
\begin{equation}\label{eq:expansion}
    \psi(t) = \sum_\alpha f_\alpha \phi_\alpha^F(t)
\end{equation}
of the fundamental solutions, the Floquet functions $\phi^F_\alpha$ 
\begin{equation}\label{eq:floquet_states}
    \phi_\alpha^F(t) = e^{-iE_\alpha t}u_\alpha(t).
\end{equation}
They are composed of the periodic function $u_\alpha$ and the non-periodic part depending on the characteristic exponent $E_\alpha$ that we  call here the Floquet energy, for reasons that will become clear shortly. It is worth noting the similarity of the Floquet functions to the Bloch functions of periodic lattices
\begin{equation}
    \phi_{\bf k}^B({\bf r}) = e^{i{\bf k r}}u_{\bf k}({\bf r}).
\end{equation}
where instead of time-periodicity we have spatial periodicity and the crystal quasi momentum plays the role of the Floquet energy. For this reason, these functions are sometimes called the Floquet-Bloch ansatz for the solution of the Schr\"odinger equations. We can take the analogy even further when we remember that the periodic part of the Bloch functions are often expanded in the basis of reciprocal lattice vectors $G$ as $u_{\bf k}({\bf r}) = \sum_{\bf G}e^{i{\bf G r}}c_{\bf k}({\bf G})$.
In the same way the time-periodic functions $u_\alpha$ can be written as
\begin{equation}\label{eq:floquet_periodic}
    u_\alpha(t) = \sum_{m=-\infty}^{\infty}e^{-im\Omega t}u^\alpha_m .
\end{equation}
This expansion into integer multiples of the fundamental Floquet frequency can be interpreted as a harmonic decomposition\cite{Shirley:1965cy} of the periodic function, where the coefficients $u^\alpha_m$ are the harmonics and we note that they generally also have a spatial dependence, i.e. $u^\alpha_m({\bf r})$. For the sum over $m$ in Eq.~(\ref{eq:floquet_periodic}) this means that it can be truncated at a finite number for the harmonics, because one does not need infinitely fast oscillating harmonics to describe physical dressing. The crystal quasi-momentum vectors $k$ are well known to be unique only within the first Brillouin zone, outside of which Bloch function are repeated with an additional $G$ phase. For Floquet functions this periodicity is in energy space and the one-dimensional Floquet-Brillouin zone has the dimension of $\Omega$. 
Despite these similarities there are however important differences between Floquet and Bloch states, because they are not eigenstates of the same kind of equation. 

\textbf{Floquet Hamiltonian}. While Bloch states solve the static Schr\"odinger equation with lattice periodic potentials Floquet states solve the time-dependent Schr\"odinger equation. Using the Floquet states of the form  Eq.~(\ref{eq:floquet_states}) together with Eq.~(\ref{eq:floquet_periodic}) as an ansatz in the time-dependent Schr\"odinger equation we get 
\begin{equation}\label{eq:HF}
    E_\alpha u^\alpha_n = \sum_{m} \left[\frac{1}{T}\int_T dt H(t)e^{i(n-m)\Omega t} -m\Omega\delta_{mn}\right]u^\alpha_m 
\end{equation}
where we have used $1/T\int_T dt\exp{[i(n-m)\Omega t]} = \delta_{mn}$. This is a matrix eigenvalue equation for the Floquet energies in the space of the harmonics
\begin{equation}
    E_\alpha u^\alpha_n = \sum_m \mathcal{H}_{nm} u^\alpha_m 
\end{equation}
and $\mathcal{H}$ is referred to as the Floquet Hamiltonian. One can see from Eq.~(\ref{eq:HF}) how the Floquet-Ansatz of developing time-dependent states into harmonics of the fundamental frequency leads to a Fourier analysis of the time-dependent Hamiltonian in terms of these harmonics. 

\textbf{Monochromatic Floquet Hamiltonian}. The decomposition of the time-dependence of the Hamiltonian into multiples of the fundamental frequency can in principle account for complex time-profiles within one period of $\Omega$. However, in practice one usually deals with monochromatic fields, in which case the Floquet Hamiltonian has a simple structure and one is allowed to discuss its general properties in detail. Expanded in the basis of the harmonic components the Floquet Hamiltonian has the matrix structure\footnote{This is only true in when the field is considered in length gauge. If one uses the velocity gauge with a vector potential ${\bf A}(t)$, as is appropriate for periodic systems, the monochromatic Hamiltonian also contains the diamagnetic term that is proportional to ${\bf A}(t)^2$.}  \begin{widetext}
\begin{equation}
   \left(
  \begin{array}{ccccccccc}
      &\vdots &\vdots &\vdots &\vdots &\vdots &\vdots &\vdots & \\
    \cdots  &  P^\dagger & H_0 - 2 \omega \mathbf{1} & P & 0 & 0 & 0 & 0 & \cdots \\
    \cdots & 0 &  P^\dagger & H_0 - \omega \mathbf{1} & P & 0 & 0 & 0 & \cdots  \\
    \cdots & 0 &  0 & P^\dagger & H_0  & P & 0 & 0  & \cdots  \\
    \cdots & 0 &  0 & 0 & P^\dagger & H_0+\omega \mathbf{1}  & P & 0  & \cdots  \\
    \cdots & 0 & 0 & 0 & 0 & P^\dagger & H_0+2 \omega \mathbf{1}  & P &  \cdots  \\
    &\vdots &\vdots &\vdots &\vdots &\vdots &\vdots &\vdots & \\
  \end{array}
  \right)
  \left(
  \begin{array}{c}
  \vdots \\ u^\alpha_{-2} \\ u^\alpha_{-1} \\ u^\alpha_{0} \\ u^\alpha_{1} \\ u^\alpha_{2} \\\vdots 
  \end{array}
  \right) = E_\alpha   \left(
  \begin{array}{c}
  \vdots \\ u^\alpha_{-2} \\ u^\alpha_{-1} \\ u^\alpha_{0} \\ u^\alpha_{1} \\ u^\alpha_2 \\\vdots 
  \end{array}
  \right)
\end{equation}
\end{widetext}
where $H_0$ is the groundstate Hamiltonian, $P=1/T\int_T dt \exp(\pm i\Omega t) H(t)$ is the expansion of the time-dependent Hamiltonian into harmonics of $\Omega$, $\mathbf{1}$ is the identity in the space of $H_0$ and $u_i$ are the harmonic components of the Floquet state.

This eigenvalue problem can equivalently expressed as a recursion relation in terms of the Floquet harmonics:
\begin{equation}\label{eq:recursion}
  (E_\alpha+m\Omega-H_0)|u^\alpha_m\rangle =P^\dagger|u^\alpha_{m-1}\rangle + P|u^\alpha_{m+1}\rangle
\end{equation}
from which one can see the periododic structure of the eigenspectrum:  each eigenvalue belongs to a set of eigenvalues shifted by multiples of $\Omega$. If two eigenvalues $E_\alpha$ and $E_{\alpha'}$ are related by multiples of $\Omega$ then their eigenstates are related as
\begin{equation}
\label{eq:shift}
E_{\alpha'}=E_{\alpha}+n\Omega \Longleftrightarrow |u^\alpha_m\rangle = |u^{\alpha'}_{m-n}\rangle    
\end{equation}
i.e. the corresponding eigenvectors differ only in the sense that they are 'shifted' with respect to the block index $m$. 

\textbf{Effective Floquet Hamiltonian} This is equivalent to the mentioned existence of a Floquet Brillouin zone in energy and one could in principle devise solution strategies focused on the energy window corresponding to this first Floquet Brillouin zone (FBZ)\cite{Sambe:1973hi}. However, the recursive structure of the Floquet eigenvalue problem also allows to define an alternative set of unique states that do not belong to the first Floquet Brillouin zone, but instead are those states that are predominantly the zero harmonics. Their eigenvalues are those that follow most closely the original bands of the problem, which is is important in many cases for the interpretation of Floquet spectra, while  one can construct the full set of eigenstates via the relation Eq.~(\ref{eq:shift}).

We thus define an effective Floquet Hamiltonian, that has the dimension of $H_0$ but that takes into account all the effects of the $n\Omega$ shifted $H_0$, such that 
\begin{equation}
  H_{eff} |u^\alpha_0\rangle = E_\alpha |u^\alpha_0\rangle  
\end{equation}
and whose eigenvalues are thus automatically zero harmonics. Starting from 
\begin{equation}
  (E_\alpha-H_0)|u^\alpha_0\rangle =P^\dagger|u^\alpha_{-1}\rangle + P|u^\alpha_{+1}\rangle
\end{equation}
and reusing the recursion we can define $H_{eff}$ as the continued fraction\cite{Perfetto:2015ila}
\begin{align}
    H_{eff}(E_\alpha) =& H_0 + P^\dagger\frac{1}{E_\alpha-\Omega -H_0 -P^\dagger \frac{1}{E_\alpha-2\Omega ...}P}P + \nonumber \\ 
    &P\frac{1}{E_\alpha+\Omega -H_0 -P \frac{1}{E_\alpha+2\Omega ...}P^\dagger}P^\dagger
\end{align}
which depends on the eigenvalue $E_\alpha$ and therefore has to be solved self-consistently. Taking the limit of large frequencies we obtain
\begin{equation}\label{eq:H_eff}
    H_{eff} = H_0 +\frac{1}{\Omega} [P^\dagger,P]
\end{equation}
which is often used to approximate the zero harmonic spectrum\cite{Mikami:2015in}.%, and c.f. section~\ref{sec:topological}.

The recursion relation \eqref{eq:recursion} also allows the interpretation of the Floquet Hamiltonian in analogy with the tight-binding formulation for electron. Writing it as
\begin{equation}
  H_0|u^\alpha_m\rangle + P^\dagger|u^\alpha_{m-1}\rangle + P|u^\alpha_{m+1}\rangle=(E_\alpha+m\Omega)|u^\alpha_m\rangle 
\end{equation}
shows how the coupling operators $P$ act as hopping terms between harmonics. 

\textbf{Floquet evolution Operator}. Floquet states and their associated eigenvalues do not only provide a useful representation of time-dependent states, but they also can be used to represent the entire time-evolution associated with the time-dependent Schr\"odinger equation. Specifically, the time-evolution operator associated with a periodic Schr\"odinger equation can be decomposed into its Floquet states:
\begin{equation}\label{eq:U}
    \hat{U}(t_1,t_2) = \sum_\alpha | \phi_\alpha^F(t_1) \rangle \langle \phi_\alpha^F(t_2)|
\end{equation}
where the sum runs again over states from the FBZ. From this formulation it is clear how knowledge of the Floquet states and eigenvalues completely solves the time-dependent problem, since given a state at any point in time, it allows the direct determination of this states at all times. One can thus express any time-dependent quantity in terms of the Floquet states in an analytical form, which we will use below for the formulation of observables in the Floquet picture.  

\textbf{Floquet analysis}. 
One can use the representation of the time evolution Eq.~(\ref{eq:U}) to verify that the expansion of the time-dependent solution of the Schr\"odinger equation into Floquet states Eq.~(\ref{eq:expansion}) is unique for all times. 
It is possible to see that the Floquet functions in the FBZ are orthogonal observing that, in 
order to satisfy the unitarity of the evolution operator,  
\begin{align}
     I &=\hat{U}^\dagger(t_1,t_2)  \hat{U}(t_1,t_2) \\
     &= \sum_{\alpha,\beta} | \phi_\alpha^F(t_2) \rangle \langle \phi_\alpha^F(t_1)| \phi_\beta^F(t_1) \rangle \langle \phi_\beta^F(t_2)|\nonumber
\end{align}
their overlap must be
\begin{equation}
    \langle\phi_\alpha^F(t)| \phi_\beta^F(t) \rangle =\delta_{\alpha,\beta}\,.
\end{equation}
Furthermore using the orthogonality condition it follows that the coefficients in the expansion of time dependent solution of Eq.~\eqref{eq:expansion} can be obtained by straightforward projection
\begin{equation}\label{eq:coefficients}
    f_\alpha=\langle\phi_\alpha^F(t)|\psi(t)\rangle\,.
\end{equation}
This result is important because it provides a direct connection between the solution of the time-dependent Schr\"odinger equation and the expansion in Floquet functions and therefore establishes a gateway to perform Floquet analysis.   
We note that such relation is independent of time provided the 
time evolution is of a pure Floquet kind. 
Any deviation from the ideal case result in an approximate validity of the expansion, yet 
Floquet analysis can still be performed provided a degree of accordance between Floquet and real time solution is achieved. 
The Floquet fidelity defined as the period averaged overlap
\begin{equation}
    F=
    \frac{1}{T}\int_\tau^{\tau+T}{\rm d} t\, |\langle\psi(t)| \sum_\alpha f_\alpha|\phi_\alpha^F(t)\rangle|^2
\end{equation}
is 1 for pure Floquet evolution and 0 otherwise and represents the most appropriate observable to quantify the degree to which Floquet analysis can be applied.

\subsection{Floquet Observables}

\textbf{Photoelectron spectrum}. Photoelectron spectroscopy provides a way to access the occupied energy levels of an electronic system.
In its general form it is governed by the matrix element 
\begin{equation}\label{eq:pe_amplitude}
    M_{\bf p}(t_f,t_i) =\langle\Psi_{\bf p}(t_f) | U(t_f,t_i) |\Psi_0(t_i) \rangle
\end{equation}
which determines the probability $P({\bf p})=w_{i\rightarrow f}=|M_{\bf p}(t_f,t_i)|^2$ to find an electron residing at a given time $t_i$ in the state $|\Psi_0(t_i)\rangle$, in the
scattering state $|\Psi_{\bf p}(t_f)\rangle$ with asymptotic momentum ${\bf p}$ at a subsequent time $t_f$ owing to the presence of an external field described by the evolution operator $U(t_f,t_i)$.
To probe the electronic structure of a Floquet system with photoelectrons one needs to consider two fields: the dressing field that creates the Floquet system and the field able to ionize the system and therefore generates a detectable ionization current. In order to see which properties of the dressed system one can access with this technique we can employ the strong field approximation (SFA)~\cite{Kitzler:2002el,Brabec:2000iz}. We further assume that the dressing field is dressing both initial and final state but ionization is taking place only by virtue of the probe laser, and that the scattering states can be described by Volkov waves.
Volkov waves are the analytical solutions of the time dependent Schr\"odinger equation of free electrons in the presence of a time dependent vector potential ${\bf A}_{pu}(t)$, and are normally expressed as plane waves, $|{\bf p}\rangle$ modulated by a time dependent phase
\begin{equation}
    |\Psi_{\bf p}^{(V)}(t)\rangle= |{\bf p}\rangle e^{-i\int_0^t{\rm d}\tau\, \frac{1}{2}({\bf p}-{\bf A}_{pu}(\tau))^2}\,.
\end{equation}
Volkov waves can be also expressed into a Floquet form assuming a monochromatic driving field~\cite{Madsen:2004kb}. Discarding the ponderomotive contribution quadratic in the field they become
\begin{equation}
    |\Psi_{\bf p}^{(V)}(t)\rangle=e^{-i \frac{{\bf p}^2}{2} t} \sum_{m=-\infty}^{\infty}e^{-i m \Omega t} |{\bf p}_m\rangle
\end{equation}
where $|{\bf p}_m\rangle=J_m(\frac{{\bf A}^0_{pu}}{\omega}\cdot {\bf p})|{\bf p}\rangle$
is a plane wave multiplied by a Bessel function whose argument depends on the dot product between the field polarization vector ${\bf A}^0_{pu}$ and the plane wave momentum ${\bf p}$.
Upon invoking these conditions the photoelectron amplitude Eq.~(\ref{eq:pe_amplitude}) becomes 
\begin{equation}\label{MSFA}
    M_{\bf p}^{(SFA)}(t_f,t_i)=-i\int_{t_i}^{t_f}{\rm d}t\, \langle\Psi_{\bf p}^{(V)}(t) | W(t) |\Psi(t) \rangle
\end{equation}
where $\Psi_{\bf p}^{(V)}(t)$ is a Volkov wave, $W(t)={\bf p}\cdot {\bf A}_{pr} +{\bf A}_{pr}^2$ describes the coupling with the probe field vector potential ${\bf A}_{pr}(t)$ and $\Psi(t)$ can be expanded with Eq.~\eqref{eq:expansion} in Floquet states dressed by the pump field ${\bf A}_{pu}(t)$.
By taking a weak monochromatic probe field ${\bf A}_{pr}(t)={\bf A}^{0}_{pr}\cos( \omega t)$ we can discard the term of the coupling quadratic in the field.
Expanding $|\Psi(t)\rangle$ and $|\Psi_{\bf p}^{(V)}(t) \rangle$ we arrive at an expression for Eq.~\eqref{MSFA} 
\begin{align}\label{MSFA1}
    M_{\bf p}^{(SFA)}&=-i \sum_\alpha \sum_{m,n} f_\alpha
    \langle {\bf p}_n|u_m^\alpha \rangle \, {\bf p}\cdot {\bf A}^0_{pr} \nonumber\\
    &\int_{-\infty}^{\infty}{\rm d}t\, e^{ i ( {\bf p}^2/2+n\Omega -E_\alpha -m\Omega-\omega ) t} \nonumber\\
    &= -i \sum_\alpha \sum_{m,n} f_\alpha
    \langle {\bf p}_n|u_m^\alpha \rangle \, {\bf p}\cdot {\bf A}^0_{pr} \nonumber\\
    &\delta( {\bf p}^2/2-E_\alpha +(n-m)\Omega-\omega)
\end{align}
where $f_\alpha$ is the Floquet expansion coefficient and we performed the time integral to obtain a delta function enforcing energy conservation.

Finally exploiting the product of delta functions we can write the photoemission probability amplitude as 
\begin{align}\label{PESSFA}
    P({\bf p})=&\sum_\alpha \sum_{n} |f_\alpha|^2 
    |u_\alpha^{n}({\bf p})|^2 ({\bf p}\cdot {\bf A}^0_{pr})^2 \nonumber\\ &\delta({\bf p}^2/2 -E_\alpha +n\Omega-\omega) \,,
\end{align}
where $u_\alpha^{n}({\bf p}) =\sum_m\langle {\bf p}_{m-n}|u_m^\alpha \rangle$.  
Based on this expression we can make some statements about which information photoelectron spectroscopy of Floquet states can extract. First of all we note that the expansion coefficients $f_{\alpha}$ enter as the square modulus in the place of the usual equilibrium band occupation. In accordance with results obtained below for the optical response function, c.f. Eq.~(\ref{eq:response_function}), we therefore interpret $|f_\alpha|^2$ as the occupation of the FBZ state. In contrast to the equilibrium case, however, this occupation does not affect only a single band, but as we can also see from Eq.~(\ref{PESSFA}) there is a series of responses at $\omega={\bf p}^2/2-E_\alpha + n \Omega$. This series of satellites are signatures of the harmonics of the Floquet states and are called Floquet sidebands. The important result here is that their intensity in a photoelectron spectrum cannot directly be interpreted as an occupation of a separate state, but instead their signal is a combination of the Floquet occupation $|f_\alpha|^2$ and the photoemission matrix elements $|u_\alpha^{n}({\bf p})|^2$ that is unique to each harmonic and decays with increasing $n$. 
Furthermore the dependence on $n$ highlights the importance of accounting for dressed continuum states in this picture as it provides different channels to enhance the probability to observe sidebands -- a mechanism known as laser assisted photoemission effect (LAPE)~\cite{Park:2014hz}.

\textbf{Optical response function} Floquet theory maps the time-dependent evolution of systems with time-dependent Hamiltonians into that of a quasi static systems as shown in Eq.~(\ref{eq:U}). The properties of such a quasi-static system can then be probed by response theory. The optical absorption spectrum can be computed from the time-dependence of the induced dipole moment of the system when perturbed by a weak but time-dependent probe field $\mathcal{E}_i(t')$
\begin{equation}\label{eq:dipole}
    d_i(t) = \int dt' \chi(t,t') \mathcal{E}_i(t')
\end{equation}
where the subscript $i$ denotes the polarization direction (which we will omit in the following for clarity) and $\chi$ is the dipole response function.

The dipole response function of any (closed) system out of equilibrium can be expressed in terms of time-evolution operators as\cite{Perfetto:2015ila}
\begin{equation}
    i\chi(t,t') = \theta(t-t')\langle \Psi(t) | \hat{d}\hat{U}(t,t')\hat{d} -h.c.|\Psi(t')\rangle
\end{equation}
where $\Psi(t)$ is the time-dependent state and $\hat{d}$ one component of the vector valued dipole operator. Specifying this general response function  to the case where the system is evolving according to Floquet time-evolution, i.e. the evolution operator is Eq.~(\ref{eq:U}) and the time dependent state $\Psi(t)$ can be expanded according to Eq.~(\ref{eq:expansion}), one obtains\cite{Perfetto:2015ila}
\begin{equation}\label{eq:response}
    i\chi(t,t') = \theta_{t-t^\prime} \sum_{\alpha\beta\gamma} f^*_\alpha f_\beta \left( d_{\alpha\gamma}(t)d_{\gamma\beta}(t^\prime) - d_{\alpha\gamma}(t^\prime) d_{\gamma\beta}(t) \right)
\end{equation}
where $\theta_{t-t^\prime}=\theta(t-t')$, and $d_{\alpha\beta}(t)=\langle \phi^F_\alpha(t)|\hat{d}|\phi^F_\beta(t)\rangle$. This expression allows for an analytical evaluation of the dipole spectrum as the Fourier transform of Eq.~(\ref{eq:dipole}) for different probe fields:
\begin{equation}\label{eq:domega}
 d(\omega') = \int dt dt^\prime e^{-i\omega' t} \chi(t,t^\prime) \mathcal{E}(t') .
\end{equation}
In experimental realizations of pump-probe optical spectroscopy, the probe is usually a short broad band pulse that can be thought of as kicking the system simultaneously at many frequencies. The theoretical idealization of such a probe is a delta function in time
\begin{equation}
    \mathcal{E}(t)=\mathcal{E}_0\delta(t-t_0) 
\end{equation}
where $t_0$ is the probe time, often referred to as the time-delay in time-resolved spectroscopies. This means that the time-dependent induced dipole-moment, Eq.~(\ref{eq:dipole}) parametrically depends on $t_0$ and we have
\begin{equation}
    d_{t_0}(t)= \int dt^\prime \chi(t,t^\prime) \mathcal{E}_0\delta(t'-t_0) \,.
\end{equation}
The Fourier transform of this expression using the Floquet dipole response function, Eq.~(\ref{eq:response}), reads
\begin{align}\nonumber
     \frac{d_{t_0}(\omega)}{\mathcal{E}_0}= 
     \sum_{\alpha\beta\gamma n m } f^*_\alpha f_\beta   d_{\alpha\gamma}^{n} d_{\gamma\beta}^{m} 
     e^{-i (\Delta E_{\beta\alpha} + (n+m)\Omega +\omega )t_0} \\
     \left(  \frac{1 }{\Delta E_{\beta\gamma } + m\Omega +\omega-i\eta}- \frac{1}{\Delta E_{\gamma\alpha }  + n\Omega +\omega-i\eta} \right)\label{eq:response_function}
\end{align}
where $d_{\alpha\beta}^{m}=\sum_{n}\langle u^{\alpha}_{n-m}|\hat{d}|u^{\beta}_{n}\rangle$ are the dipole matrix elements between harmonic components of FBZ states and  $\Delta E_{\beta\alpha } = E_\beta-E_\alpha$. This expression is useful because its pole structure shows the features of the optical response: unless suppressed by matrix elements or the expansion coefficients the optical response of a Floquet system has poles corresponding to transitions between the satellite bands. This shows that while a Floquet electronic structure is established in a material, the eigenmodes of the system are the Floquet quasi-energies, even if the probe is instantaneous and it is not necessary to average over the pump cycle. The Floquet states are instantaneously observable. The probe time $t_0$ enters here by determining a complex phase that affects the lineshape of resonances in a non-trivial way.

It is worth pointing out that the Floquet response function can be written as
\begin{align}\nonumber
     \frac{d_{t_0}(\omega)}{\mathcal{E}_0}=& 
     \sum_{\alpha\beta n m } (|f_\alpha|^2- |f_\beta|^2)   \frac{d_{\alpha\beta}^{n} d_{\beta\alpha}^{m} 
     e^{-i ((n+m)\Omega +\omega )t_0}} {\Delta E_{\alpha\beta } + m\Omega +\omega-i\eta}+\\\nonumber
     &\sum_{\alpha\neq\beta,\gamma n m } f^*_\alpha f_\beta   d_{\alpha\gamma}^{n} d_{\gamma\beta}^{m} 
     e^{-i (\Delta E_{\beta\alpha} + (n+m)\Omega +\omega )t_0} \\
     &\left(  \frac{1 }{\Delta E_{\beta\gamma } + m\Omega +\omega-i\eta}- \frac{1}{\Delta E_{\gamma\alpha }  + n\Omega +\omega-i\eta} \right)\label{eq:response_function2}
\end{align}
and the first term closely resembles the standard Lehman representation of equilibrium response functions. It allows for the interpretation of the squared modulus of the Floquet expansion coefficients $|f_\alpha|^2$ as occupations of FBZ states in the sense of equilibrium distribution functions, that affect the absorption spectrum via Pauli blocking. We note however, that the second term is generally non-zero and indeed accounts for all response features when $|f_\alpha|^2=|f_\beta|^2$. 

In the spirit of Floquet analysis, the response function derived above is not intended as a prescription to compute experimental spectra. As we will see below, the experimental reality of non-equilibrium systems is more complex than the Floquet picture. Instead, this response functions represent the ideal limit of a pump-probe probe setup where the pump has established a stable Floquet electronic structure and the probe is infinitely short. It can thus be used to guide the interpretation of experiments even if the conditions are not perfectly reached.

\textbf{High harmonic generation}. 
Floquet theory can be naturally used to describe non-linear response phenomena 
such as high harmonic generation (HHG). HHG occurs when a system, pumped by a 
strong monochromatic field, emits radiation at higher frequencies. 
The spectrum of the emitted radiation is governed by the Larmor formula
\begin{equation}\label{eq:hhglarmor}
S(\omega)\propto \left|\mathscr{F}\left[\frac{\partial i(t)}{\partial t}\right](\omega)\right|^2=\omega^2|i(\omega)|^2 \end{equation}
that describes the radiation emitted by an accelerated charge density in terms of the Fourier transform of the time derivative of the current, $i(t)$.
Below we present a Floquet formulation of HHG that involves only bound states, and is therefore applicable to solids where ionization is not the dominant generation mechanism. In order to describe HHG in atoms and molecules one needs to include continuum states in the Floquet expansion, see e.g.~\onlinecite{Kapoor:2012jp}.

The time-dependent current ${\bf i}(t)$ is the expectation value of the current density operator $\hat{\bf j}$ over the time-dependent state $\psi(t)$
\begin{align}\label{eq:hhgit}
    {\bf i}(t) &=\langle \psi(t) |\hat{\bf j}|\psi(t)\rangle \\
            &=\frac{1}{2}\left[ \langle \psi(t) |\hat{\bf p}|\psi(t)\rangle  -\langle \psi(t)^* |\hat{\bf p}|\psi(t)^*\rangle -2 \frac{A(t)}{c} |\psi(t)|^2 \right]\nonumber\\
            &=  \langle \psi(t) |\hat{\bf j}_0|\psi(t)\rangle - \frac{A(t)}{c} |\psi(t)|^2
\end{align}
where $\hat{\bf j}_0$ is the paramagnetic current operator and the vector potential (in the velocity gauge), $A(t)$, accounts for the diamagnetic component.

By expanding $\psi(t)$ in Floquet states and taking the Fourier transform of~\eqref{eq:hhgit} one obtains 
\begin{equation}
    {\bf i}(\omega)=\sum_{\alpha \beta n} f_\alpha^*f_\beta {\bf j}_{\alpha \beta}^n
    \delta(\Delta E_{\beta \alpha} + n\Omega +\omega) - \frac{A(\omega)}{c} N
\end{equation}
where ${\bf j}_{\alpha \beta}^n=\sum_m\langle u^\alpha_{m-n} |\hat{\bf j}_0|u^\beta_m\rangle$ and $|\psi(t)|^2=N$ since the norm of the wavefunction is a
constant equal to the number of electrons. This result shows how the Floquet expansion can give insight into complex non-linear spectra, such as HHG, where a simple interpretation in terms of band electrons is no longer possible. Instead the Floquet picture provides an interpretation of the spectral features in terms of 'transitions' between Floquet states and their associated matrix elements.

\section{Floquet analysis demonstrated}
In this section we will demonstrate the concepts of Floquet analysis with a simple illustrative model. First we show how Rabi oscillation is linked to the Floquet picture. Then we discuss the Floquet coefficients of the Floquet expansion of the solutions of the time-dependent Schr\"odinger equation that form the link between actual time evolutions and Floquet states. The interpretative capability of the Floquet picture is shown for the example of photoelectron spectroscopy of this simple model and finally we demonstrate how the linear dipole response functions is able to perfectly describe the non-equilibrium response of this simple model.

We consider a driven two level system
\begin{equation}\label{eq:Rabi}
  H(t) = \left(\begin{array}{cc}
      \frac{\epsilon}{2} & 0  \\
      0 & -\frac{\epsilon}{2}  
  \end{array} \right)  + 
  A(t) \cos(\Omega t)\left(\begin{array}{cc}
      0 & M  \\
      M^* & 0  
  \end{array} \right)
\end{equation}
where $\epsilon$ is the energy difference of the levels, $M$ the coupling matrix element between the levels induced by the pump and $A(t)$ the amplitude of the pump field. To solve this model explicitly we evaluate $\psi(t)=\exp(-i\int_0^t d\tau H(\tau))|\psi(0)$, where $\psi(0)=(0,1)$. We set $\epsilon=1$, so that $\Omega=1$ is the resonant excitation case, and let $M=0.05$. 

{\bf Rabi oscillation}. With constant amplitude $A(t)=A_0$ equation~(\ref{eq:Rabi}) is the paradigm for Rabi oscillations where the overlap of solutions of the time-dependent Schr\"odinger equation with the groundstate, $|\langle\psi_0|\psi(t) \rangle|^2$, oscillates with the Rabi frequency $\Omega_{Rabi}=\sqrt{(A_0M)^2 +(\epsilon-\Omega)^2}$. This oscillation only indirectly depends on the fundamental frequency of the drive through the detuning from the level spacing $(\epsilon-\Omega)$ and on resonance is directly proportional to the coupling and field strength. This means, solutions of the time-dependent Schr\"odinger equation oscillate with a frequency that is independent of the fundamental frequency, yet it is intimately related to the Floquet solution. Diagonalizing the Floquet Hamiltonian associated with Eq.~(\ref{eq:Rabi}) yields for the Floquet Brillouin zone eigenvalues
\begin{equation}\label{eq:floquet_eigenvalues}
    E_{1,2} = \frac{1}{2}(\Omega\pm\sqrt{(A_0M)^2 +(\epsilon-\Omega)^2})
\end{equation}
under the approximation of weak coupling.%(c.f. Appendix).%<--Same as in dressing Nanoletter 
These Floquet eigenvalues determine the non-periodic phase in the Floquet ansatz, Eq.~(\ref{eq:floquet_states}), and  their combination determines the phase of the time-dependent state, Eq.~(\ref{eq:expansion}). Specifically, taking the square modulus of the overlap with the groundstate
\begin{align}\nonumber
|\langle\psi_0|\psi(t) \rangle|^2 =& |f_1|^2|\langle\psi_0|u_1(t) \rangle|^2+|f_2|^2|\langle\psi_0|u_2(t) \rangle|^2+ \\
& f_1^*f_2e^{i(E_2-E_1)t}\langle u_1(t)|\psi_0 \rangle\langle\psi_0|u_2(t) \rangle+ c.c.
\label{eq:Rabi_osc}
\end{align}
one sees that it oscillates with $E_2- E_1=\Omega_{Rabi}$. The Rabi frequency is thus the difference of the FBZ eigenvalues, which is also referred to Rabi splitting. This simple result shows how the time-resolved projection that is in atomic physics interpreted as occupation transfer and the hybridization of static Floquet sidebands are two ways of looking at the same fundamental process. The difference only appears in the observation: to observe Rabi oscillation of the occupation one needs to refer to the groundstate, which is an eigenstate of the static Hamiltonian and as such generally not part of the solutions of the time-dependent Schr\"odinger equation where the pump field is present. Therefore, to measure the occupation with photoelectron spectroscopy or optical spectroscopy one needs to switch the field off. The Floquet picture, instead, gives the spectrum of the system while the field is switched on and in order to observe the Rabi splitting one needs to probe the system together with the pump.

We also note that the Rabi frequency is not the only time-dependence in Eq.~(\ref{eq:Rabi_osc}) and that other terms can contribute multiples of the fundamental frequency which can lead to beatings between the Rabi frequency and the fundamental mode. 

\begin{figure*}
    \centering
    \includegraphics[width=0.8\textwidth]{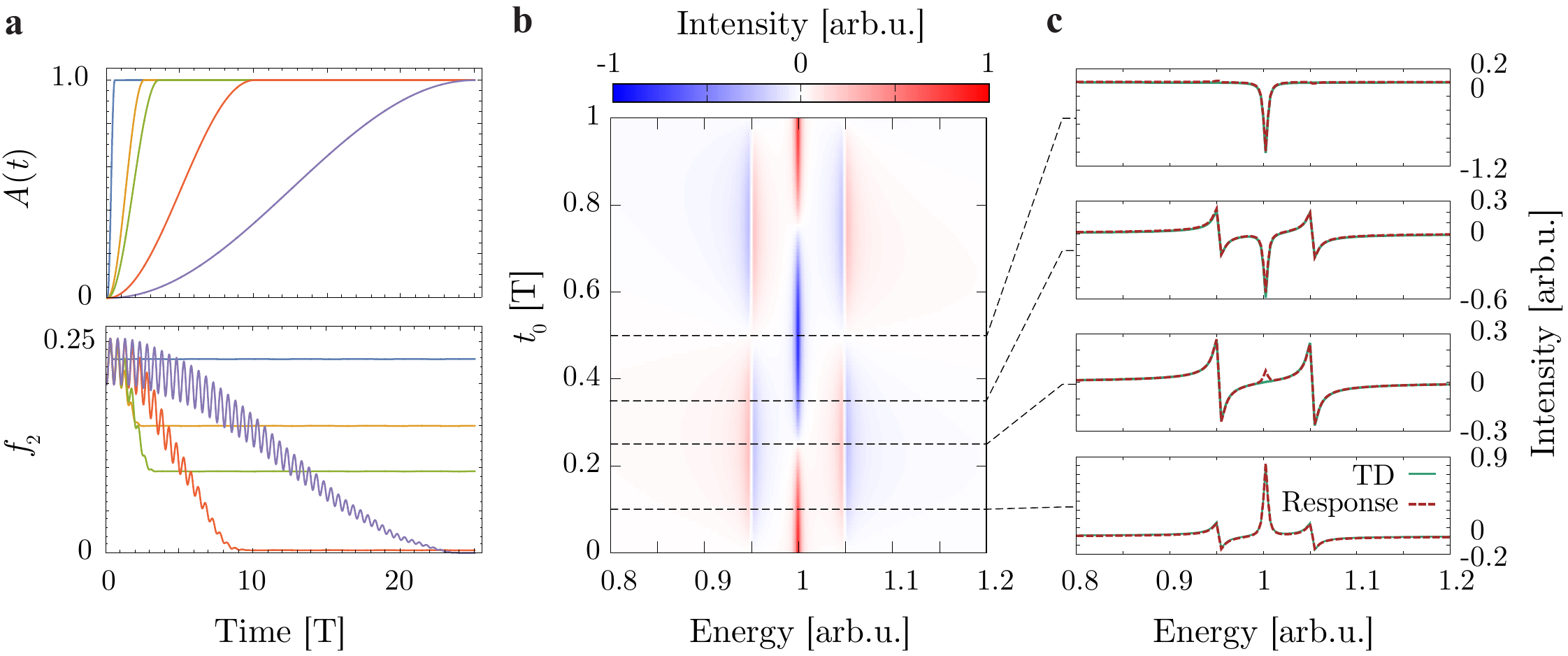}
    \caption{{\bf  Floquet analysis of a simple model}: ({\bf a}) Floquet expansion of a time-evolving solution for the Schr\"odinger equation for different pump envelopes with an off-resonant energy, $\Omega=0.8$. The upper panel shows a series of $A(t)$ that smoothly reach a constant plateau at different times (measured in units of pump cycle $T$). The lower panel shows the numerically computed expansion coefficient $f_2$ for the upper Floquet state. One can see that when the pump envelope is constant the Floquet expansion is valid. Very slow switch-on times lead to almost no contribution of the Floquet state to the time evolution and the coefficient vanishes.  ({\bf b}) Dipole response of the resonantly driven two level system for various probe time $t_0$ throughout one cycle $T$ of the drive frequency. Depending on the probe time satellites corresponding to transitions between replica bands are emerging around the main resonance at 1 [a.u.]. The effect of the probe time  induced complex phase in the response is to completely reverse the sign of the spectrum when scanned though the cycle, but the spectrum is periodic after one cycle. ({\bf c}) Different cuts through the scan of ({\bf b}) at given probe times, show the comparison between the explicit time-evolution results (red dashed lines) and the evaluated Floquet response functions Eq.~(\ref{eq:response_function}), (solid blue lines).     }  
    \label{fig:model}
\end{figure*}

{\bf Expansion coefficients}. In the above example we assumed a constant field amplitude, but in an experiment one needs to start from the groundstate and turn on the pump field to reach the Floquet regime. During the switch-on phase Floquet theory does not apply to the time-dependent Schr\"odinger equation, but we can still compute the expansion of a numerical time-dependent solutions with varying switch on profiles into Floquet states, Eq.~(\ref{eq:expansion}). We numerically compute the time-dependent solution of this model and compute the coefficients $f_\alpha$ according to Eq.~(\ref{eq:coefficients}) at all times.  We can observe how they become time-independent once the pulse shape reaches a plateau, Fig.~\ref{fig:model}a, and the system is in the Floquet regime. For the two level system on resonance the time-dependent state is described by an equal contribution of both FBZ states, i.e. $|f_1|=|f_2|$, and the system always reaches this configuration independently of the switch-on time.  

The independence of the coefficients on the evolution of the system only occurs for perfect resonance in the two level system. When we allow for an imbalance between the states by considering off-resonance, the system has the freedom to adjust its configuration and the situation is drastically different. As shown in Fig.~\ref{fig:model}a, the switch on time determines completely the Floquet configuration.  The state with the smaller coefficient is a Floquet state that resembles more the second level and can be interpreted as a replica of the excited state. The instantaneous quench provides an upper bound for the coefficient of this state and with increasing switch-on time the coefficient approaches zero, while it approaches one for the other state. This means for slow switch-on the time-dependent state retains mainly the character of the ground state. Invoking the interpretation of occupation for Floquet coefficients, this means that only the lowest Floquet state has occupation. This interpretation is supported by Eq.~(\ref{eq:Rabi_osc}) where no Rabi oscillation takes place when one of the coefficients vanishes. Still, the absence of Rabi oscillation does not imply the absence of Rabi splitting, as the eigenvalue of the occupied Floquet state is still $E_1$ from Eq.~(\ref{eq:floquet_eigenvalues}) and is shifted due to the presence of the pump. This means that while only one Floquet state is occupied, there is still a finite occupation in the excited state and if one would switch off the field this excited state population would be exposed. By the same reasoning perfect resonance behaviour always presents Rabi oscillation irrespective of the switch on. 

{\bf Photoelectron spectroscopy}.
The photoelectron spectrum of a two level system can be thought of as the Fourier transform of the projection of the time-dependent state to the occupied initial state
\begin{equation}
    P(\omega) = \left|\frac{1}{\sqrt{2\pi}}\int_{-\infty}^\infty dt e^{-i\omega t}\langle \psi(0)|\psi(t)\rangle \right|^2 .
\end{equation}
This equation can be evaluated either numerically from the explicit time evolution of the Schr\"odinger equation, or for Floquet systems by inserting the Floquet expansion, Eq.~(\ref{eq:expansion}), for the time dependent state, giving
\begin{equation}\label{eq:model_PES}
    P(\omega) = \left|\frac{1}{\sqrt{2\pi}} \sum_{\alpha,m} f_\alpha\langle\psi(0)|u^\alpha_m\rangle\delta(E_\alpha+m\Omega-\omega) \right|^2 .
\end{equation}
Since the simple two level system that we are considering here lacks spatial resolution and therefore has no momentum, this is equivalent to the Floquet-photoelectron expression given in Eq.~(\ref{PESSFA}) and we can discuss some features of photoelectron spectroscopy of Floquet systems by considering this simplified example. 

First we note that, as observed above, the spectrum consists of a series of peaks centered around each FBZ energy level, $E_\alpha$, and  spaced by multiples of the pump frequency. These peaks, however, can only be observed if two conditions are met: The FBZ state $\psi^F_\alpha(t)$ needs to have a finite contribution to the time evolving solution of the Schr\"odinger equation, i.e. $f_\alpha$ has to be non-zero, and the harmonic components of the Floquet state $u^\alpha_m$ need to have a finite overlap with the occupied groundstate. Especially the last condition is crucial for the observation of sidepeaks and is governed by the intensity of pump and the strength of the coupling.     

The Floquet expression of pump-probe photoelectron also illustrates the important fact noted above in the context of Rabi oscillation: The observable energy levels of a pumped system are the Floquet eigenvalues and from Eq.~(\ref{eq:model_PES}) we see that there is no reference to the energy levels of the equilibrium system. While the Floquet expansion coefficients $f_\alpha$ depend on the switch-on of the pump, the Floquet energy levels only depend on its intensity and the coupling strength. This means that slow or fast switch-on conditions will affect the intensity of certain peaks in the spectrum, they will not however, affect the energy levels themselves.  

{\bf Optical response}. 
Similar to the case of photoelectron spectroscopy Floquet analysis can also guide the interpretation of pump-probe optical spectroscopy via the Floquet dipole response function Eq.~(\ref{eq:response_function}). We demonstrate here how the non-equilibrium response function reproduces results from explicit time-evolution of the driven two level system, Eq.~(\ref{eq:Rabi}). From the explicit numerical time evolution of the Schr\"odinger equation we obtain the dipole response by applying an additional "kick" potential $V_{\rm probe}(t) = \delta(t-t_0)$ to probe the system and by then evaluating the induced dipole $d(t)=\langle \psi(t)| \hat{d}|\psi(t)\rangle -\langle \psi_0(t)| \hat{d}|\psi_0(t)\rangle$, where $\psi_0$ is the evolution of the reference state without a kick and $\hat{d}=\sigma_x$ is a Pauli matrix. The frequency dependent induced dipole response equivalent to Eq.~(\ref{eq:response_function}) is then obtained by Fourier transform 
\begin{equation}
    d_{t_0}(\omega) =\frac{1}{\sqrt{2\pi}} \int_{-\infty}^\infty dt e^{-i\omega t}e^{-\eta\theta(t-t_0)(t-t_0) }d(t),    
\end{equation}
where we have added the broadening factor $\eta$, c.f. Refs.~\onlinecite{DeGiovannini:2013dr, Walkenhorst:2016hc,DeGiovannini:2018bl}. The results for a resonant pump frequency and with different kick delays $t_0$ is shown in Fig.~\ref{fig:model}b and we observe that the delays significantly affect the line shape of the response and that sidepeaks originating from transition between Rabi-split levels are enhanced or suppressed by different delays. 
 
Most importantly Fig.~\ref{fig:model}c also shows the comparison to the analytical Floquet-response function for different delays and the agreement is excellent. Here we are considering the resonant case and hence the Floquet coefficients are $|f_1|=|f_2|$ as noted above. This means that the first term in Eq.~(\ref{eq:response_function2}), that resembles the equilibrium response function, is vanishing and all observable response is generated by the terms containing $\alpha\neq\beta$. This shows that the response treatment of Floquet systems is a not a simple extension of the equilibrium case and that it describes non-trivial non-equilibrium responses. 

Here, we have only shown the resonant case, where the effect of the switch-on phase of the pump on the Floquet expansion coefficients $f_\alpha$ plays no role, but we note that in more general off-resonant cases this leads to a further rich modification of the spectra.
 
\section{Photon driven Floquet systems}
\textbf{Topological Floquet Materials}
Among most studied classes of Floquet matter are topological materials. Generally speaking, these are theoretical proposals, where the photon dressed electronic states of solid state systems have a different topological nature than the equilibrium states. This idea has generated considerable interested because it provides a route towards controlling and tuning topological properties in materials on a fast time-scale and in a reversible fashion, thus making the various features associated with topological edge states potentially amenable for a variety of technological applications, such as for example computing or metrology.

One can classify the proposed topological Floquet materials in two broad categories. One uses the Floquet picture of the electronic structure to show that the application of an electromagnetic field changes the topology of the system or modifies its topological features. Such proposals usually consider host materials that have non-trivial topology in equilibrium and in many cases use the effective Hamiltonian Eq.~(\ref{eq:H_eff}) that relies on the high-frequency approximation suitable for semi-metals. The other of the two broad classes comprises proposals that aim to turn materials that have trivial topology in equilibrium into Floquet-topological materials. This usually relies on the Floquet-replica mechanism where equilibrium bands of different character are brought into (near) resonance such that new hybridizations of electronic states are created.   

The first works proposing this idea have concentrated on two-dimensional systems, specifically the ac-field based realization of the Haldane model for a Chern insulator in real materials\cite{oka_photovoltaic_2009,Inoue:2010iz, Kitagawa:2011dr}. While Haldane's original proposal\cite{haldane_model_1988} relies on a static magnetic field to break time-reversal symmetry and thus endow the Dirac states of a hexagonal system with a non-trivial mass term and hence a Chern number, these works propose to use circularly polarized lasers to create a dressed electronic structure that is described by a Floquet Hamiltonian with the same topological properties. In particular the work by Oka \textit{et. al}\cite{oka_photovoltaic_2009} focuses on the experimental observable of such an induced change in the topology. They predict a finite and quantized Hall current, in line with the notion that the Floquet system is a Chern insulator and hence hosts the anomalous Hall effect. 
Following these early works based on graphene a large variety of proposal's have emerged that aim at manipulating the topology of this 2D Dirac system\cite{Dora:2012jd,Grushin:2014gt,PerezPiskunow:2014iy,Usaj:2014bl,Dahlhaus:2015iy,Mikami:2015in,Sentef:2015jp,Dutreix:2016ck,Qu:2018dc} and surface states of topological insulators\cite{Fregoso:2013di}. With the recent discovery of higher dimensional Dirac and Weyl materials\cite{Armitage:2018dg}, this line of research has considerably broadened. For these materials especially, the high-frequency expansion of the Floquet Hamiltonian, Eq.~(\ref{eq:H_eff}) has proven to be very prolific, because it affords an analytic expression for the dressed Hamiltonian and as such can be readily classified in terms of topology. Thus, there have been proposals for the manipulation of Floquet-topological phases in line-node semimetals \cite{Taguchi:2016ho,Narayan:2016jl,Ezawa:2017gv}, Dirac-semimetals\cite{Saha:2016cz,Chan:2016dqa,Ebihara:2016de,Hubener:2017ht} and Weyl-semimetals  \cite{Chan:2016ir,Yan:2016eea,Yan:2017bv,Zhang:2016di,Taguchi:2016ef} all relying on the effective Hamiltonian description.

Still using the effective Hamiltonian formulation, all such Floquet-topological phases have been systematically classified\cite{Yao:2017fk}.
It has also been used to propose the Floquet topological Magnon\cite{Owerre:2017fj,Owerre:2018eo,Owerre:2018dz}.

The above examples all have in common that the equilibrium system already has non-trivial topology, or at least as in the case of graphene, is at a phase boundary in a topological phase diagram. It is, however, also possible to turn topological trivial systems into topological ones by applying circularly polarized light. This was first proposed for 2D quantum wells where the Floquet quasi-energy structure was demonstrated to feature the edge states of topological insulators\cite{lindner_floquet_2011}. The mechanism by which this is achieved is the inversion of the band character across the band gap that is the hallmark of topological insulators and is here realized by the Floquet mechanism: The continuous driving on resonance with the bandgap creates replica states of the valence band at the energy of the conduction bands, leading to a hybridization of the states and an inversion of the band character of the Floquet states. This opens the perspective of inducing topological properties and in particular protected edge currents into otherwise trivial materials only by applying lasers. Similar proposals have explored this pathway of designing topological materials properties\cite{Katan:2013hl,sie_valley-selective_2014,Claassen:2016ge,Liu:2018dk}. 

\begin{figure*}
    \centering
    \includegraphics[width=0.9\textwidth]{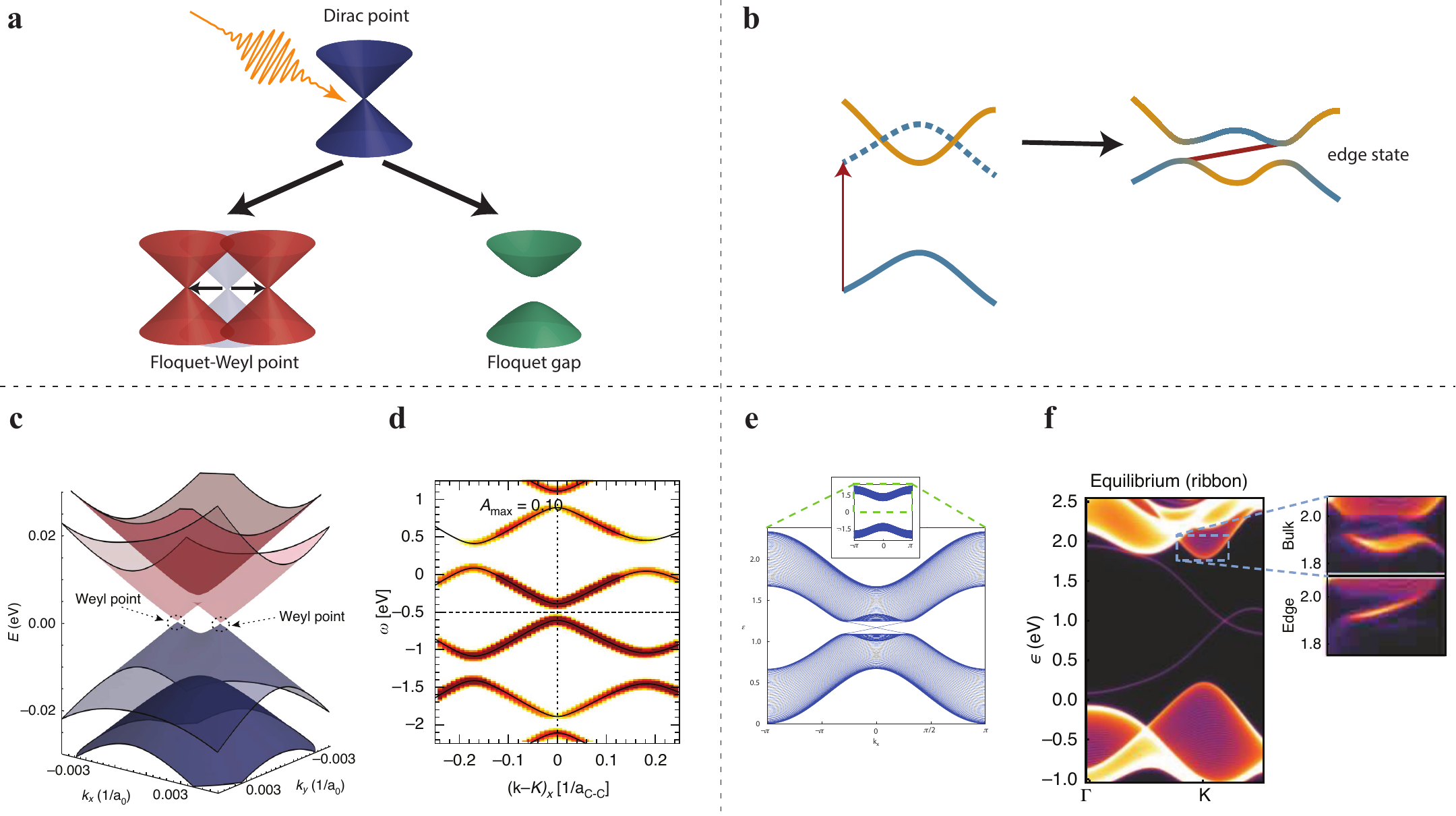}
    \caption{{\bf Proposed Floquet topological materials}: ({\bf a}) High frequency circularly polarized pumping of Dirac bands leads to Floquet-topological phase transitions, e.g. 3D Dirac semimetal to Floquet-Weyl semimetal (left) and 2D Dirac semimetal into a Floquet-Chern insulator (right). ({\bf b}) (Near)-Resonant pumping of trivial semiconductors leads to hybridization of Floquet bands that host protected edge states. 
    ({\bf c}) TDDFT-Floquet-bandstructure of bulk Na$_3$Bi showing two Floquet-Weyl points originating from a single 3D Dirac point. ({\bf d}) Computed ARPES spectrum of circularly polarized pumped graphene showing the opening of a gap at the Dirac point. ({\bf e}) Computed bandstructure of a Floquet topological insulator from 2D quantum wells in a striped geometry showing hybridization of Floquet replica bands and the Floquet topological protected edge state. ({\bf f}) Computed Floquet spectrum of a pumped stripe of WS$_2$ monolayer, showing hybridization and protected edge state. Adapted from Ref.~\onlinecite{Hubener:2017ht}, Macmillan Publishers Ltd ({\bf c}); Ref.~\onlinecite{Sentef:2015jp}, Macmillan Publishers Ltd ({\bf d}); Ref.~\onlinecite{lindner_floquet_2011}, Macmillan Publishers Ltd ({\bf e}); Ref.~\onlinecite{Claassen:2016ge}, Macmillan Publishers Ltd ({\bf f})}
    \label{fig:topology}
\end{figure*}

\begin{figure*}
    \centering
    \includegraphics[width=0.9\textwidth]{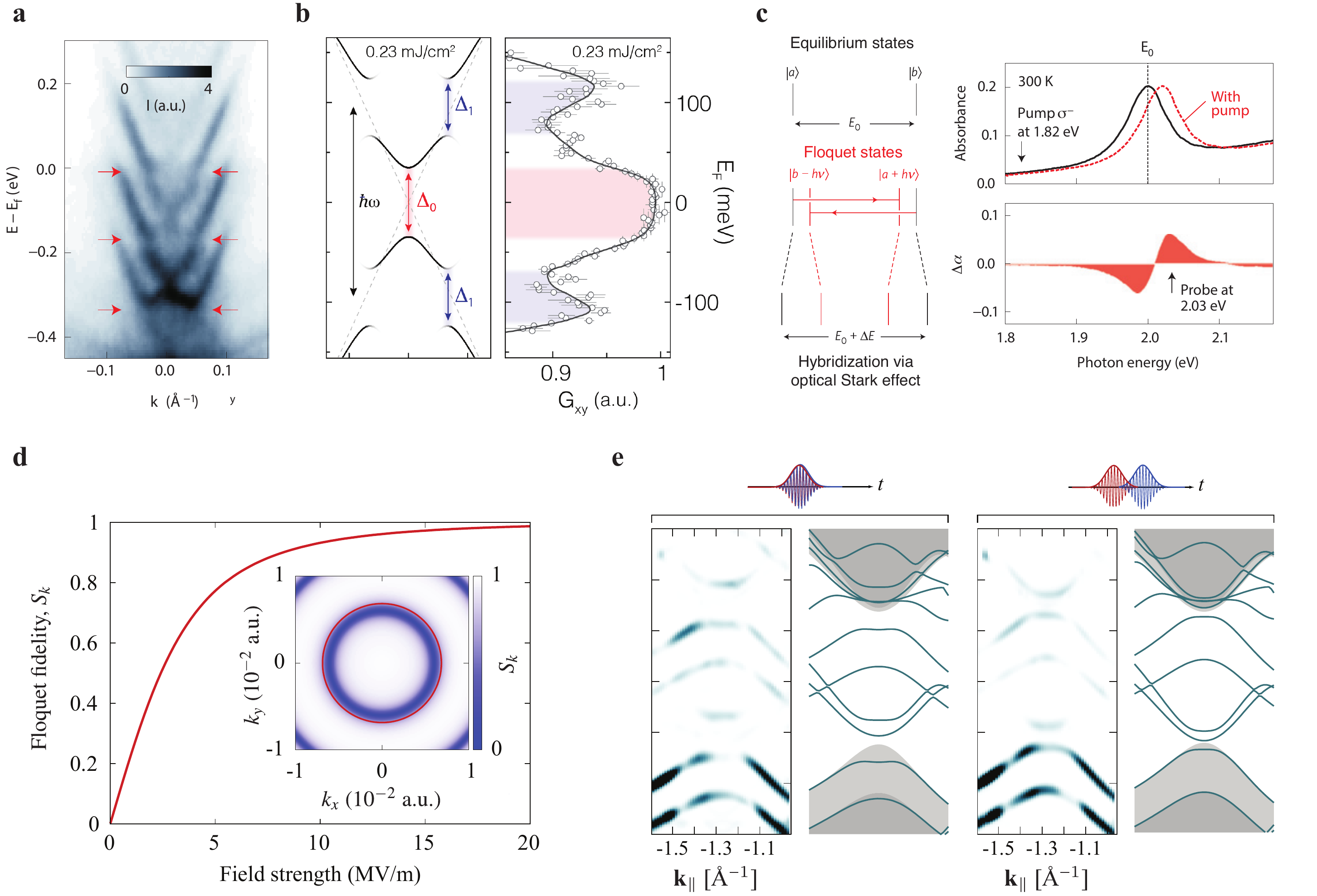}
    \caption{{\bf  Observation of photon driven Floquet-phases}: ({\bf a}) ARPES measurement showing Floquet replicas of Dirac bands at the surface of the topological insulator Bi$_2$Se$_3$ and small opening of gaps indicated by red arrows. ({\bf c}) Measured doping dependent Hall conductance (right panel) of Floquet bands (left panel) in graphene. As the Fermi level is scanned across an energy region corresponding to the pump frequency, clear transport signatures are observable in correspondence to the computed gap openings of Floquet bands. ({\bf c}) Valley selective optical Stark effect in WS$_2$, where Floquet bands hybridize leading to a shift of the optical response and are proposed to host a topological edge state.  ({\bf c}) Floquet analysis of the experiment shown in  ({\bf b}) by considering a dissipative Dirac model and computing how well dynamical states are described by Floquet theory, quantified as Floquet fidelity. It confirms that Floquet states are well established throughout the BZ (inset) except at the resonant gap (red circle) and shows that strong fields are required to establish a stable Floquet phase with good fidelity.  ({\bf e}) Computed ARPES spectra and their Floquet analysis of pumped WSe$_2$ showing the formation of Floquet bands for different pump-probe overlaps (top row cartoons). The photo-electron spectra are very well matched with the Floquet bands, giving the underlying band picture to the observed Stark effect shown in ({\bf c}). Adapted from Ref.~\onlinecite{Mahmood:2016bu}, Macmillan Publishers Ltd ({\bf a}); Ref.~\onlinecite{2018McIver}, with permission from the authors ({\bf b}); Ref.~\onlinecite{sie_valley-selective_2014}, Macmillan Publishers Ltd ({\bf c}); Ref.~\onlinecite{Sato2019a}, with permission from the authors ({\bf d}); Ref.~\onlinecite{deGiovannini:2016cb}, ACS ({\bf e}).}  
    \label{fig:photon}
\end{figure*}

{\bf Experimental Observation of Floquet topological phases}
The first unambiguous observation of topological Floquet bands have been reported in Refs.~\onlinecite{Wang:2013fe,Mahmood:2016bu} as photoelectron spectra of the topological insulator Bi$_2$Se$_3$. The Dirac bands that form at the surface of this material have here been shown to develop hybridization gaps under irradiation with light, shown in Fig.~\ref{fig:photon}a. The opening of hybridization gaps is the most widely predicted feature of Floquet topological phases,
%, c.f. Sec.~\ref{sec:topological},
however, as pointed out in Refs.~\onlinecite{Park:2014hz,Mahmood:2016bu} in an photo-emission experiment the external pump laser can also dress electronic states outside the material. Such a dressing, known as the light-assisted photo-emission (LAPE) effect, also results in a series of satellite bands, but does not reflect a dressing of the electronic structure of the material. Therefore photo-electron experiments have to be carefully designed to account for this dressing effect.

Besides the opening of a gap for Dirac bands, Floquet theory also predicts hybridization gaps from crossing of replica bands and some works propose that this interaction induces topological properties in topological trivial materials. In particular in transition metal dichalcogenides, it is expected\cite{sie_valley-selective_2014,Claassen:2016ge} that a band inversion of Floquet bands induces topological edge states. While such bands have not been directly observed, Ref.~\onlinecite{sie_valley-selective_2014} report a detailed study of a valley selective Stark effect in theses materials, that can be directly interpreted in terms of Floquet analysis, Fig.~\ref{fig:photon}c. The observed valley-dependent circular dichroism results from an interaction of Floquet-replica bands that opens the band gap in either of the valleys, depending on the orientation of the pump polarization. In the optical spectroscopy of Ref.~\onlinecite{sie_valley-selective_2014} this is detected as a Stark shift, while Ref.~\onlinecite{deGiovannini:2016cb} reports the Floquet analysis of simulated photo-electron spectra of this system. The computed angular-resolved photoelectron spectroscopy (ARPES) probabilities are shown for different pump-probe delays Fig.~\ref{fig:photon}e and directly compared to the Floquet spectrum. The results show the hybridization of the sidebands underlying the reported Stark effect, confirmed by Floquet analysis. A noteworthy detail of this work is, that the Floquet analysis agrees with the ARPES spectra even if pump and probe are not perfectly overlapping, which in principle violates the basic assumption of Floquet theory of perfect time-translation invariance. Instead, this shows that even for finite pulse setups the Floquet picture is applicable and can be used for interpretation of experimental results.

While the observation of Floquet bands is still being pursued, experimental confirmation of topological properties of Floquet phases is also underway. Topological materials in equilibrium host an anomalous Hall current, i.e. a Hall current that flows as soon as a source drain voltage is applied without requiring a magnetic field. The origin of this current is a purely quantum mechanical effect, because it is directly proportional the Chern number of the material, i.e. the integrated Berry curvature\cite{Thouless:1982kq}. For Floquet topological systems it is therefore expected that the Berry curvature of Floquet states results in the same kind of Hall current.

The challenge in observing this current in a pump-probe set up is that it has to be detected at very fast time-scales, because one needs to use ultrafast pump pulses to avoid damaging the material. In Ref.~\onlinecite{2018McIver} an ultrafast transport measurement setup based on optical switches is presented that achieves to detect current with a resolution of ~1ps. With this setup the authors succeed in obtaining the light induced  Hall conductivity in graphene. The dependence of the Hall conductivity on gate-doping of the graphene sample, shown in Fig.~\ref{fig:photon}b reveals that currents are generated from the modified Floquet bandstructure, confirming that the Floquet Dirac bands have opened hybridization gaps in accordance with the predictions based on Floquet theory.       

In equilibrium materials the origin of an anomalous Hall current can be unambiguously related the the Berry curvature structure, however in driven systems a careful analysis of the excitation process is required to relate an observed Hall current to the topological nature of the Floquet bands. Using the concept of Floquet fidelity, a measure of how well a given dynamical system is described by its corresponding Floquet states, the authors of Ref.~\onlinecite{Sato2019a} show that for the strong pump pulse intensities used in the experiment of Ref.~\onlinecite{2018McIver}, Floquet states are well established throughout the Brillouin zone (BZ) as shown in Fig.~\ref{fig:photon}d. The observed Hall current, however, is found to result not purely from the topology of the Floquet states, but is partly due to an imbalance in the population of excited states created by the pump.     

The proposals for Floquet phases presented thus far, all use external radiation to realize the dressed electronic structure. However, as shown in the experiment reported in Ref.~\onlinecite{2018McIver} and pointed out in many theoretical works\cite{Dehghani:2014jm,Dehghani:2015gz,Moessner:2017jb} continued application of a laser to solids results in heating and eventually damage of the sample. Therefore, in experiments a delicate balance needs to be struck between the time necessary to establish the Floquet phase and the maximum time the material can sustain the radiation. In this context it is important to note that dissipative systems, where the energy can redistributed to a heat bath can help to stabilize the Floquet phase\cite{Moessner:2017jb,Sato2019a,Sato2019b}. In the next section, instead, we will explore the possibility to achieve establishing Floquet dressed states without continued transfer of energy into the system.

\section{Phonon dressed Floquet matter}
The basic idea underlying the realization of Floquet matter is the existence of a time periodic potential in the Hamiltonian. The origin of this potential does not need to be an externally applied laser, but instead can be provided by an internal mode of the system. Here, we will discuss the concept of Floquet matter created from dressing with phonons\cite{Murakami:2017ht,Hubener:2018id}. A somewhat similar idea, but without invoking a Floquet dressing picture was proposed earlier\cite{Garate:2013er} and the concept of electron-phonon coupling induced topological phase transitions has gained some traction\cite{Wang:2017by,Antonius:2016gj}.

Especially in photo-electron spectroscopy, features originating from strong electron-phonon coupling are well known. The best known example being the electron-phonon kinks in the ARPES spectrum of superconducting materials\cite{Lanzara:2001kp}. However, electron phonon coupling is also known to result in observable distinct satellite, or shadow,  bands in the photo-electron spectral function\cite{Lee:2014bw,Antonius:2015fj,Story:2014hn}, also known as Polaron replicas\cite{Moser:2013bf,Chen:2015fz,Cancellieri:2016fw,Wang:2016de,Verdi:2017bx}. The avenue to create a phonon-driven Floquet material is to selectively excite such strongly coupled phonon modes and thus to create new properties of the electronic structure or to better understand its properties as for example realized in Ref.~\onlinecite{Gerber:2017bm}. The crucial distinction to photon driven Floquet states is, that the systems needs to be excited first externally, but the dressing then is provided by an eigenmode of the system. As such the Floquet phonon picture is more of a framework to look at internal interactions of excitations in a material in the sense of Floquet analysis, rather than providing a shortcut to computing non-equilibrium properties, as it is, instead, often done with the photon dressing picture. 

\begin{figure*}
    \centering
    \includegraphics[width=1.0\textwidth]{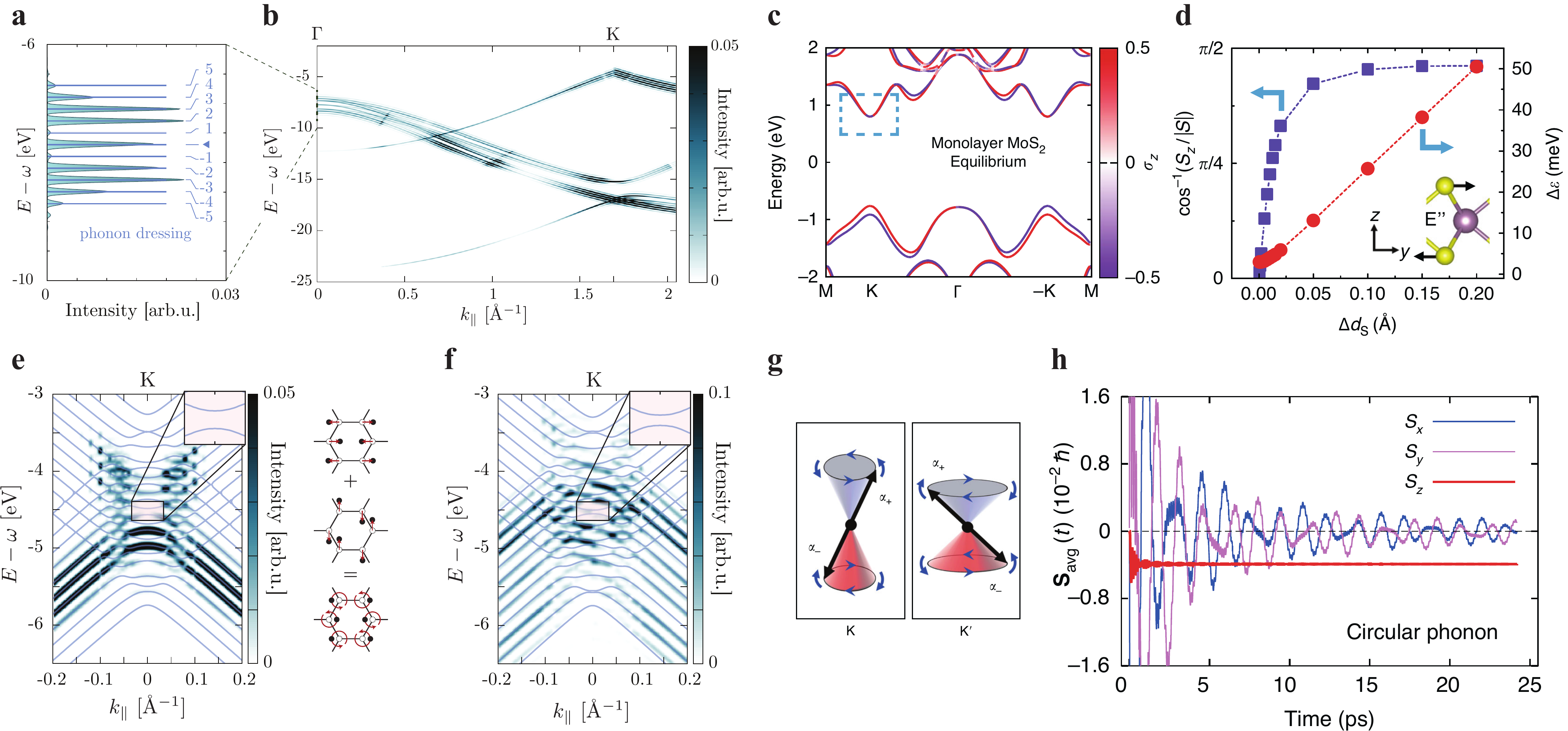}
    \caption{{\bf Observables of Floquet-Phonon Matter}: ({\bf a}) Computed ARPES spectral function and Floquet analysis of graphene pumped by its E$_2g$ mode at the $\Gamma$ point of the BZ. ({\bf b}) Computed ARPES spectrum like in ({\bf a}) but along a path through the full BZ. ({c}) Bandstructure and spin-polarization of monolayer MoS$_2$ that is strongly modified by the $E''$ phonon mode at K. ({\bf d}) Change in spin polarization of the conduction band at $K$ as a function of lattice displacement along the $E''$ phonon mode. ({\bf e}) computed ARPES spectrum like in ({\bf a}) Computed ARPES spectral function and Floquet analysis of graphene pumped by a circularly polarized E$_2g$ mode (c.f. cartoon on the right) around the $K$ point of the BZ. The inset shows an enlargement of the Floquet-phonon bands with an open gap at the Dirac point. ({\bf f}) Same as ({\bf e}) but the dressing field is a circularly polarized laser, showing the equivalence of photon and phonon dressing. ({\bf g}) Spin component of circular-phonon dressed Floquet states at the conduction band bottom at the $K$ and $K'$ points of the BZ, showing a difference in the magnitude of $S_z$ component. ({\bf h}) This results in an overall finite spin polarization (red line) in time of the full material. Adapted from Ref.~\onlinecite{Hubener:2018id}, ACS ({\bf a}),({\bf b}),({\bf e}) and ({\bf f}); Ref.~\onlinecite{Shin:2018ek}, Macmillan Publishers Ltd ({\bf c}), ({\bf d}), ({\bf g}) and ({\bf h}). 
    }
    \label{fig:phonon}
\end{figure*}

As an example for Floquet-phonon matter we consider here the work of Ref.~\onlinecite{Hubener:2018id}, which presents a detailed comparison between the computed photo-electron spectrum of phonon dressed graphene and the Floquet analysis of this dressing. The optical E$_2g$ phonon mode is dynamically realized by perturbing the lattice and propagating numerically the coupled time-dependent density functional theory (TDDFT) and Ehrenfest equations. This is equivalent to a state of the material where this mode has been excited by an external field as a coherent lattice oscillation and then this field has been switched off. In the photo-electron spectrum this internal excitation manifests as a series of driven polaronic satellites, c.f. Fig.~\ref{fig:phonon}a, that, however vary strongly across the electronic BZ, c.f. Fig.~\ref{fig:phonon}b. Performing the Floquet analysis of the same time-dependent Hamiltonian, yields the complete energy bands of these satellite series as shown for the spectral function at the $\Gamma$ point, Fig.~\ref{fig:phonon}a. The Floquet quasi-energy levels are, as expected, equally spaced replicas across the full BZ, in contrast to strong variation in the ARPES spectral function. As mentioned before, 
% we need to mention this somewhere before!!!
the intensity of the Floquet sidebands depends generally on the amplitude of the drive and the strength of the coupling. Since, the electrons at all $k$-points feel the same coherent phonon, this variation reflects the strong variation of the (dynamical) electron-phonon coupling across the BZ.

The phonon-dressing of the electronic structure can also be used to affect the topological properties of the material, in a similar way as has been proposed for photon dressing. This requires a time-reversal symmetry breaking driving mode, which in the case for phonons can be achieved by exciting degenerate modes coherently such that the atoms perform a circular trajectory around their equilibrium position. The E$_2g$ mode in graphene consist of longitudinal and transverse branches that are degenerate at the phonon $\Gamma$-point, so that exciting them with a relative phase delay of $\pi/2$ achieves  the rotating motion. Figure~\ref{fig:phonon}e shows the ARPES spectrum of such an excited system around Dirac point superimposed with its Floquet analysis. In analogy to the proposals using photon drives, Fig.~\ref{fig:phonon}f, the circular polarized phonon induces a non-trivial gap opening at the Dirac point and the material has undergone a topological phase transition.        

Another example where the phonon-dressing of the electronic structure results in changed materials properties is the phonon driven Floquet-magnetization presented in Ref.~\onlinecite{Shin:2018ek}. Monolayer transition metal dichalcogenides can be excited selectively in either of their non-equivalent $K$-valleys\cite{sie_valley-selective_2014} by circularly polarized light. The spin of such an excitation is strongly coupled to the $E''$ optical phonon mode, where even a relatively small lattice displacement induces large changes in the spin polarization of the conduction band bottom as shown in Fig.~\ref{fig:phonon}c\&d. Inducing the phonon as a circularly polarized mode, as in Ref.~\onlinecite{Hubener:2018id}, results in a phonon dressed electronic state of this conduction band, the Floquet phonon states, that turn out to have precessing spin polarization and thus have induced local magnetization, c.f. Fig.~\ref{fig:phonon}g. Such a circular polarized phonon 
breaks time reversal symmetry, with the consequence that the spin-precession in both non-equivalent $K$ valleys do not have the same magnitude, which results in an overall magnetization, as shown in Fig.~\ref{fig:phonon}h, of this otherwise non-magnetic material. This is one of the few examples of how the Floquet-phonon dressed electronic structure can yield radically different material properties than its host material in equilibrium. This kind of phonon induced magnetization has been reported for a different material in Ref.~\cite{Nova:2017ja}

\section{Perspective}
We have shown how new phases and properties of materials under non-equilibrium can be understood in terms of Floquet theory. Especially the observation of phenomena directly originating from the dressing of the electronic structure is very encouraging for the realizations of targeted design of materials through the Floquet mechanism. In this brief review we have focused on the connection between the theoretical construct of Floquet theory and experimentally observable signatures. We have shown how Floquet theory can be used as a tool of analysis of excitation mechanisms and towards interpretation of experiments. One example of such interpretation that we have only mentioned briefly here but deserves more attention could be high-harmonic generation in solids, which itself is the detection of harmonics in the optical response and as such it should be very worthwhile approaching it with the Floquet analysis paradigm. 

Most proposals to establish Floquet phases in materials use external lasers as the source for the dressing field. Here we have discussed the possibility of using internal eigenmodes of the material to create a dressing field and have demonstrated how this kind of dressed electronic structure emerges from phonon dressing. One can, however, envision a variety of other modes. In particular plasmon modes of the electronic structure are well studied in terms of an expansion of so called cumulants\cite{Guzzo:2011dc,Guzzo:2014gu,Kas:2014dj} in the plasma frequency and are understood to result in plasmon polaron satellites in photo-electron spectra\cite{Offi:2007hq,Guzzo:2014gu}, even resulting in replica bandstructures\cite{Caruso:2015fj,Lischner:2015fd}. Hence, we expect that targeted excitation of such modes will lead to rich dressing physics.

\section{Acknowledgements}
The authors are grateful for discussions with A. Rubio, S. A. Sato, X. Liu, P. Tang, D. Shin, N. Park and M. Sentef. This work was supported by the European Research Council (ERC-2015-AdG694097).

% \bibliography{bib}

\begin{thebibliography}{10}
\expandafter\ifx\csname url\endcsname\relax
  \def\url#1{\texttt{#1}}\fi
\expandafter\ifx\csname urlprefix\endcsname\relax\def\urlprefix{URL }\fi
\providecommand{\bibinfo}[2]{#2}
\providecommand{\eprint}[2][]{\url{#2}}

\bibitem{Hsieh:2017ix}
\bibinfo{author}{Hsieh, D.}, \bibinfo{author}{Basov, D.~N.} \&
  \bibinfo{author}{Averitt, R.~D.}
\newblock \bibinfo{title}{{Towards properties on demand in quantum materials}}.
\newblock \emph{\bibinfo{journal}{Nature Materials}}
  \textbf{\bibinfo{volume}{16}}, \bibinfo{pages}{1077--1088}
  (\bibinfo{year}{2017}).

\bibitem{Mitrano:2016fr}
\bibinfo{author}{Mitrano, M.} \emph{et~al.}
\newblock \bibinfo{title}{{Possible light-induced superconductivity in K3C60 at
  high temperature}}.
\newblock \emph{\bibinfo{journal}{Nature}} \textbf{\bibinfo{volume}{530}},
  \bibinfo{pages}{461--464} (\bibinfo{year}{2016}).

\bibitem{Mankowsky:2014em}
\bibinfo{author}{Mankowsky, R.} \emph{et~al.}
\newblock \bibinfo{title}{{Nonlinear lattice dynamics as a basis for enhanced
  superconductivity in YBa2Cu3O6.5}}.
\newblock \emph{\bibinfo{journal}{Nature}} \textbf{\bibinfo{volume}{516}},
  \bibinfo{pages}{71--73} (\bibinfo{year}{2014}).

\bibitem{Stojchevska:2014it}
\bibinfo{author}{Stojchevska, L.} \emph{et~al.}
\newblock \bibinfo{title}{{Ultrafast Switching to a Stable Hidden Quantum State
  in an Electronic Crystal}}.
\newblock \emph{\bibinfo{journal}{Science}} \textbf{\bibinfo{volume}{344}},
  \bibinfo{pages}{177--180} (\bibinfo{year}{2014}).

\bibitem{Nova:2017ja}
\bibinfo{author}{Nova, T.~F.} \emph{et~al.}
\newblock \bibinfo{title}{{An effective magnetic field from optically driven
  phonons}}.
\newblock \emph{\bibinfo{journal}{Nature Physics}}
  \textbf{\bibinfo{volume}{13}}, \bibinfo{pages}{132--136}
  (\bibinfo{year}{2017}).

\bibitem{Aidelsburger:2018jf}
\bibinfo{author}{Aidelsburger, M.}, \bibinfo{author}{Nascimbene, S.} \&
  \bibinfo{author}{Goldman, N.}
\newblock \bibinfo{title}{{Artificial gauge fields in materials and engineered
  systems}}.
\newblock \emph{\bibinfo{journal}{Comptes Rendus Physique}}
  \textbf{\bibinfo{volume}{19}}, \bibinfo{pages}{394--432}
  (\bibinfo{year}{2018}).

\bibitem{Oka:2018df}
\bibinfo{author}{Oka, T.} \& \bibinfo{author}{Kitamura, S.}
\newblock \bibinfo{title}{{Floquet Engineering of Quantum Materials}}.
\newblock \emph{\bibinfo{journal}{Annual Review of Condensed Matter Physics}}
  \textbf{\bibinfo{volume}{10}},
  \bibinfo{pages}{annurev--conmatphys--031218--013423} (\bibinfo{year}{2018}).

\bibitem{Eckardt:2017hc}
\bibinfo{author}{Eckardt, A.}
\newblock \bibinfo{title}{{<i>Colloquium</i>: Atomic quantum gases in
  periodically driven optical lattices}}.
\newblock \emph{\bibinfo{journal}{Rev. Mod. Phys.}}
  \textbf{\bibinfo{volume}{89}}, \bibinfo{pages}{011004}
  (\bibinfo{year}{2017}).

\bibitem{Rechtsman:2013fe}
\bibinfo{author}{Rechtsman, M.~C.} \emph{et~al.}
\newblock \bibinfo{title}{{Photonic Floquet topological insulators}}.
\newblock \emph{\bibinfo{journal}{Nature}} \textbf{\bibinfo{volume}{496}},
  \bibinfo{pages}{196--200} (\bibinfo{year}{2013}).

\bibitem{Jotzu:2014kz}
\bibinfo{author}{Jotzu, G.} \emph{et~al.}
\newblock \bibinfo{title}{{Experimental realization of the topological Haldane
  model with ultracold fermions}}.
\newblock \emph{\bibinfo{journal}{Nature}} \textbf{\bibinfo{volume}{515}},
  \bibinfo{pages}{237--240} (\bibinfo{year}{2014}).

\bibitem{Wang:2018fy}
\bibinfo{author}{Wang, Y.} \emph{et~al.}
\newblock \bibinfo{title}{{Theoretical understanding of photon spectroscopies
  in correlated materials in and out of equilibrium}}.
\newblock \emph{\bibinfo{journal}{Nature Reviews Materials}}
  \textbf{\bibinfo{volume}{3}}, \bibinfo{pages}{312--323}
  (\bibinfo{year}{2018}).

\bibitem{Eisert:2015ka}
\bibinfo{author}{Eisert, J.}, \bibinfo{author}{Friesdorf, M.} \&
  \bibinfo{author}{Gogolin, C.}
\newblock \bibinfo{title}{{Quantum many-body systems out of equilibrium}}.
\newblock \emph{\bibinfo{journal}{Nature Physics}}
  \textbf{\bibinfo{volume}{11}}, \bibinfo{pages}{124--130}
  (\bibinfo{year}{2015}).

\bibitem{Moessner:2017jb}
\bibinfo{author}{Moessner, R.} \& \bibinfo{author}{Sondhi, S.~L.}
\newblock \bibinfo{title}{{Equilibration and order in quantum Floquet matter}}.
\newblock \emph{\bibinfo{journal}{Nature Physics}}
  \textbf{\bibinfo{volume}{13}}, \bibinfo{pages}{424--428}
  (\bibinfo{year}{2017}).

\bibitem{Floquet:1883eo}
\bibinfo{author}{Floquet, G.}
\newblock \bibinfo{title}{{Sur les {\'e}quations diff{\'e}rentielles
  lin{\'e}aires {\`a} coefficients p{\'e}riodiques}}.
\newblock \emph{\bibinfo{journal}{Annales scientifiques de l'{\'E}cole Normale
  Sup{\'e}rieure}} \textbf{\bibinfo{volume}{12}}, \bibinfo{pages}{47--88}
  (\bibinfo{year}{1883}).

\bibitem{Shirley:1965cy}
\bibinfo{author}{Shirley, J.~H.}
\newblock \bibinfo{title}{{Solution of the Schr{\"o}dinger Equation with a
  Hamiltonian Periodic in Time}}.
\newblock \emph{\bibinfo{journal}{Physical Review}}
  \textbf{\bibinfo{volume}{138}}, \bibinfo{pages}{B979--B987}
  (\bibinfo{year}{1965}).

\bibitem{Note1}
\bibinfo{note}{This is only true in when the field is considered in length
  gauge. If one uses the velocity gauge with a vector potential ${\protect \bf
  A}(t)$, as is appropriate for periodic systems, the monochromatic Hamiltonian
  also contains the diamagnetic term that is proportional to ${\protect \bf
  A}(t)^2$.}

\bibitem{Sambe:1973hi}
\bibinfo{author}{Sambe, H.}
\newblock \bibinfo{title}{{Steady States and Quasienergies of a
  Quantum-Mechanical System in an Oscillating Field}}.
\newblock \emph{\bibinfo{journal}{Physical Review A}}
  \textbf{\bibinfo{volume}{7}}, \bibinfo{pages}{2203--2213}
  (\bibinfo{year}{1973}).

\bibitem{Perfetto:2015ila}
\bibinfo{author}{Perfetto, E.} \& \bibinfo{author}{Stefanucci, G.}
\newblock \bibinfo{title}{{Some exact properties of the nonequilibrium response
  function for transient photoabsorption}}.
\newblock \emph{\bibinfo{journal}{Physical Review A}}
  \textbf{\bibinfo{volume}{91}}, \bibinfo{pages}{033416}
  (\bibinfo{year}{2015}).

\bibitem{Mikami:2015in}
\bibinfo{author}{Mikami, T.} \emph{et~al.}
\newblock \bibinfo{title}{{Brillouin-Wigner theory for high-frequency expansion
  in periodically driven systems: Application to Floquet topological
  insulators}}.
\newblock \emph{\bibinfo{journal}{Physical Review B}} \bibinfo{pages}{144307}
  (\bibinfo{year}{2015}).
\newblock \eprint{1511.00755}.

\bibitem{Kitzler:2002el}
\bibinfo{author}{Kitzler, M.}, \bibinfo{author}{Milosevic, N.},
  \bibinfo{author}{Scrinzi, A.}, \bibinfo{author}{Krausz, F.} \&
  \bibinfo{author}{Brabec, T.}
\newblock \bibinfo{title}{{Quantum Theory of Attosecond XUV Pulse Measurement
  by Laser Dressed Photoionization}}.
\newblock \emph{\bibinfo{journal}{Physical Review Letters}}
  \textbf{\bibinfo{volume}{88}}, \bibinfo{pages}{173904}
  (\bibinfo{year}{2002}).

\bibitem{Brabec:2000iz}
\bibinfo{author}{Brabec, T.} \& \bibinfo{author}{Krausz, F.}
\newblock \bibinfo{title}{{Intense few-cycle laser fields: Frontiers of
  nonlinear optics}}.
\newblock \emph{\bibinfo{journal}{Reviews Of Modern Physics}}
  \textbf{\bibinfo{volume}{72}}, \bibinfo{pages}{545--591}
  (\bibinfo{year}{2000}).

\bibitem{Madsen:2004kb}
\bibinfo{author}{Madsen, L.~B.}
\newblock \bibinfo{title}{{Strong-field approximation in laser-assisted
  dynamics}}.
\newblock \emph{\bibinfo{journal}{American Journal of Physics}}
  \textbf{\bibinfo{volume}{73}}, \bibinfo{pages}{57--62}
  (\bibinfo{year}{2004}).

\bibitem{Park:2014hz}
\bibinfo{author}{Park, S.~T.}
\newblock \bibinfo{title}{{Interference in Floquet-Volkov transitions}}.
\newblock \emph{\bibinfo{journal}{Physical Review A}}
  \textbf{\bibinfo{volume}{90}}, \bibinfo{pages}{013420}
  (\bibinfo{year}{2014}).

\bibitem{Kapoor:2012jp}
\bibinfo{author}{Kapoor, V.} \& \bibinfo{author}{Bauer, D.}
\newblock \bibinfo{title}{{Floquet analysis of real-time wave functions without
  solving the Floquet equation}}.
\newblock \emph{\bibinfo{journal}{Physical Review A}}
  \textbf{\bibinfo{volume}{85}}, \bibinfo{pages}{023407}
  (\bibinfo{year}{2012}).

\bibitem{DeGiovannini:2013dr}
\bibinfo{author}{De~Giovannini, U.} \emph{et~al.}
\newblock \bibinfo{title}{{Simulating Pump-Probe Photoelectron and Absorption
  Spectroscopy on the Attosecond Timescale with Time-Dependent Density
  Functional Theory}}.
\newblock \emph{\bibinfo{journal}{Chemphyschem}} \textbf{\bibinfo{volume}{14}},
  \bibinfo{pages}{1363--1376} (\bibinfo{year}{2013}).

\bibitem{Walkenhorst:2016hc}
\bibinfo{author}{Walkenhorst, J.}, \bibinfo{author}{De~Giovannini, U.},
  \bibinfo{author}{Castro, A.} \& \bibinfo{author}{Rubio, A.}
\newblock \bibinfo{title}{{Tailored pump-probe transient spectroscopy with
  time-dependent density-functional theory: controlling absorption spectra}}.
\newblock \emph{\bibinfo{journal}{The European Physical Journal B}}
  \textbf{\bibinfo{volume}{89}}, \bibinfo{pages}{128} (\bibinfo{year}{2016}).

\bibitem{DeGiovannini:2018bl}
\bibinfo{author}{De~Giovannini, U.} \& \bibinfo{author}{Castro, A.}
\newblock \bibinfo{title}{{CHAPTER 12:Real-time and Real-space Time-dependent
  Density-functional Theory Approach to Attosecond Dynamics}}.
\newblock In \emph{\bibinfo{booktitle}{Attosecond Molecular Dynamics}},
  \bibinfo{pages}{424--461} (\bibinfo{publisher}{Royal Society of Chemistry},
  \bibinfo{address}{Cambridge}, \bibinfo{year}{2018}).

\bibitem{oka_photovoltaic_2009}
\bibinfo{author}{Oka, T.} \& \bibinfo{author}{Aoki, H.}
\newblock \bibinfo{title}{{Photovoltaic Hall effect in graphene}}.
\newblock \emph{\bibinfo{journal}{Physical Review B}}
  \textbf{\bibinfo{volume}{79}}, \bibinfo{pages}{081406}
  (\bibinfo{year}{2009}).

\bibitem{Inoue:2010iz}
\bibinfo{author}{Inoue, J.-i.} \& \bibinfo{author}{Tanaka, A.}
\newblock \bibinfo{title}{{Photoinduced Transition between Conventional and
  Topological Insulators in Two-Dimensional Electronic Systems}}.
\newblock \emph{\bibinfo{journal}{Physical Review Letters}}
  \textbf{\bibinfo{volume}{105}}, \bibinfo{pages}{017401}
  (\bibinfo{year}{2010}).

\bibitem{Kitagawa:2011dr}
\bibinfo{author}{Kitagawa, T.}, \bibinfo{author}{Oka, T.},
  \bibinfo{author}{Brataas, A.}, \bibinfo{author}{Fu, L.} \&
  \bibinfo{author}{Demler, E.}
\newblock \bibinfo{title}{{Transport properties of nonequilibrium systems under
  the application of light: Photoinduced quantum Hall insulators without Landau
  levels}}.
\newblock \emph{\bibinfo{journal}{Physical Review B}}
  \textbf{\bibinfo{volume}{84}}, \bibinfo{pages}{235108}
  (\bibinfo{year}{2011}).

\bibitem{haldane_model_1988}
\bibinfo{author}{Haldane, F. D.~M.}
\newblock \bibinfo{title}{{Model for a Quantum Hall Effect without Landau
  Levels: Condensed-Matter Realization of the "Parity Anomaly"}}.
\newblock \emph{\bibinfo{journal}{Physical Review Letters}}
  \textbf{\bibinfo{volume}{61}}, \bibinfo{pages}{2015--2018}
  (\bibinfo{year}{1988}).

\bibitem{Dora:2012jd}
\bibinfo{author}{D{\'o}ra, B.}, \bibinfo{author}{Cayssol, J.},
  \bibinfo{author}{Simon, F.} \& \bibinfo{author}{Moessner, R.}
\newblock \bibinfo{title}{{Optically Engineering the Topological Properties of
  a Spin Hall Insulator}}.
\newblock \emph{\bibinfo{journal}{Physical Review Letters}}
  \textbf{\bibinfo{volume}{108}}, \bibinfo{pages}{056602}
  (\bibinfo{year}{2012}).

\bibitem{Grushin:2014gt}
\bibinfo{author}{Grushin, A.~G.}, \bibinfo{author}{G{\'o}mez-Le{\'o}n, {\'A}.}
  \& \bibinfo{author}{Neupert, T.}
\newblock \bibinfo{title}{{Floquet Fractional Chern Insulators}}.
\newblock \emph{\bibinfo{journal}{Physical Review Letters}}
  \textbf{\bibinfo{volume}{112}}, \bibinfo{pages}{156801}
  (\bibinfo{year}{2014}).

\bibitem{PerezPiskunow:2014iy}
\bibinfo{author}{Perez-Piskunow, P.~M.}, \bibinfo{author}{Usaj, G.},
  \bibinfo{author}{Balseiro, C.~A.} \& \bibinfo{author}{Torres, L. E. F.~F.}
\newblock \bibinfo{title}{{Floquet chiral edge states in graphene}}.
\newblock \emph{\bibinfo{journal}{Physical Review B}}
  \textbf{\bibinfo{volume}{89}}, \bibinfo{pages}{121401}
  (\bibinfo{year}{2014}).

\bibitem{Usaj:2014bl}
\bibinfo{author}{Usaj, G.}, \bibinfo{author}{Perez-Piskunow, P.~M.},
  \bibinfo{author}{Torres, L. E. F.~F.} \& \bibinfo{author}{Balseiro, C.~A.}
\newblock \bibinfo{title}{{Irradiated graphene as a tunable Floquet topological
  insulator}}.
\newblock \emph{\bibinfo{journal}{Physical Review B}}
  \textbf{\bibinfo{volume}{90}}, \bibinfo{pages}{115423}
  (\bibinfo{year}{2014}).

\bibitem{Dahlhaus:2015iy}
\bibinfo{author}{Dahlhaus, J.~P.}, \bibinfo{author}{Fregoso, B.~M.} \&
  \bibinfo{author}{Moore, J.~E.}
\newblock \bibinfo{title}{{Magnetization Signatures of Light-Induced Quantum
  Hall Edge States}}.
\newblock \emph{\bibinfo{journal}{Physical Review Letters}}
  \textbf{\bibinfo{volume}{114}}, \bibinfo{pages}{246802}
  (\bibinfo{year}{2015}).

\bibitem{Sentef:2015jp}
\bibinfo{author}{Sentef, M.~A.} \emph{et~al.}
\newblock \bibinfo{title}{{Theory of Floquet band formation and local
  pseudospin textures in pump-probe photoemission of graphene}}.
\newblock \emph{\bibinfo{journal}{Nature Communications}}
  \textbf{\bibinfo{volume}{6}}, \bibinfo{pages}{7047} (\bibinfo{year}{2015}).

\bibitem{Dutreix:2016ck}
\bibinfo{author}{Dutreix, C.}, \bibinfo{author}{Stepanov, E.~A.} \&
  \bibinfo{author}{Katsnelson, M.~I.}
\newblock \bibinfo{title}{{Laser-induced topological transitions in phosphorene
  with inversion symmetry}}.
\newblock \emph{\bibinfo{journal}{Physical Review B}}
  \textbf{\bibinfo{volume}{93}}, \bibinfo{pages}{241404}
  (\bibinfo{year}{2016}).

\bibitem{Qu:2018dc}
\bibinfo{author}{Qu, C.}, \bibinfo{author}{Zhang, C.} \&
  \bibinfo{author}{Zhang, F.}
\newblock \bibinfo{title}{{Valley-selective topologically ordered states in
  irradiated bilayer graphene}}.
\newblock \emph{\bibinfo{journal}{2D Materials}} \textbf{\bibinfo{volume}{5}},
  \bibinfo{pages}{011005} (\bibinfo{year}{2018}).

\bibitem{Fregoso:2013di}
\bibinfo{author}{Fregoso, B.~M.}, \bibinfo{author}{Wang, Y.~H.},
  \bibinfo{author}{Gedik, N.} \& \bibinfo{author}{Galitski, V.}
\newblock \bibinfo{title}{{Driven electronic states at the surface of a
  topological insulator}}.
\newblock \emph{\bibinfo{journal}{Physical Review B}}
  \textbf{\bibinfo{volume}{88}}, \bibinfo{pages}{155129}
  (\bibinfo{year}{2013}).

\bibitem{Armitage:2018dg}
\bibinfo{author}{Armitage, N.~P.}, \bibinfo{author}{Mele, E.~J.} \&
  \bibinfo{author}{Vishwanath, A.}
\newblock \bibinfo{title}{{Weyl and Dirac semimetals in three-dimensional
  solids}}.
\newblock \emph{\bibinfo{journal}{Rev. Mod. Phys.}}
  \textbf{\bibinfo{volume}{90}}, \bibinfo{pages}{015001}
  (\bibinfo{year}{2018}).

\bibitem{Taguchi:2016ho}
\bibinfo{author}{Taguchi, K.}, \bibinfo{author}{Xu, D.-H.},
  \bibinfo{author}{Yamakage, A.} \& \bibinfo{author}{Law, K.~T.}
\newblock \bibinfo{title}{{Photovoltaic anomalous Hall effect in line-node
  semimetals}}.
\newblock \emph{\bibinfo{journal}{Physical Review B}}
  \textbf{\bibinfo{volume}{94}}, \bibinfo{pages}{155206}
  (\bibinfo{year}{2016}).

\bibitem{Narayan:2016jl}
\bibinfo{author}{Narayan, A.}
\newblock \bibinfo{title}{{Tunable point nodes from line-node semimetals via
  application of light}}.
\newblock \emph{\bibinfo{journal}{Physical Review B}}
  \textbf{\bibinfo{volume}{94}}, \bibinfo{pages}{041409}
  (\bibinfo{year}{2016}).

\bibitem{Ezawa:2017gv}
\bibinfo{author}{Ezawa, M.}
\newblock \bibinfo{title}{{Photoinduced topological phase transition from a
  crossing-line nodal semimetal to a multiple-Weyl semimetal}}.
\newblock \emph{\bibinfo{journal}{Physical Review B}}
  \textbf{\bibinfo{volume}{96}}, \bibinfo{pages}{041205}
  (\bibinfo{year}{2017}).

\bibitem{Saha:2016cz}
\bibinfo{author}{Saha, K.}
\newblock \bibinfo{title}{{Photoinduced Chern insulating states in semi-Dirac
  materials}}.
\newblock \emph{\bibinfo{journal}{Physical Review B}}
  \textbf{\bibinfo{volume}{94}}, \bibinfo{pages}{081103}
  (\bibinfo{year}{2016}).

\bibitem{Chan:2016dqa}
\bibinfo{author}{Chan, C.-K.}, \bibinfo{author}{Oh, Y.-T.},
  \bibinfo{author}{Han, J.~H.} \& \bibinfo{author}{Lee, P.~A.}
\newblock \bibinfo{title}{{Type-II Weyl cone transitions in driven
  semimetals}}.
\newblock \emph{\bibinfo{journal}{Physical Review B}}
  \textbf{\bibinfo{volume}{94}}, \bibinfo{pages}{121106}
  (\bibinfo{year}{2016}).

\bibitem{Ebihara:2016de}
\bibinfo{author}{Ebihara, S.}, \bibinfo{author}{Fukushima, K.} \&
  \bibinfo{author}{Oka, T.}
\newblock \bibinfo{title}{{Chiral pumping effect induced by rotating electric
  fields}}.
\newblock \emph{\bibinfo{journal}{Physical Review B}}
  \textbf{\bibinfo{volume}{93}}, \bibinfo{pages}{155107}
  (\bibinfo{year}{2016}).

\bibitem{Hubener:2017ht}
\bibinfo{author}{H{\"u}bener, H.}, \bibinfo{author}{Sentef, M.~A.},
  \bibinfo{author}{de~Giovannini, U.}, \bibinfo{author}{Kemper, A.~F.} \&
  \bibinfo{author}{Rubio, A.}
\newblock \bibinfo{title}{{Creating stable Floquet-Weyl semimetals by
  laser-driving of 3D Dirac materials}}.
\newblock \emph{\bibinfo{journal}{Nature Communications}}
  \textbf{\bibinfo{volume}{8}}, \bibinfo{pages}{13940} (\bibinfo{year}{2017}).

\bibitem{Chan:2016ir}
\bibinfo{author}{Chan, C.-K.}, \bibinfo{author}{Lee, P.~A.},
  \bibinfo{author}{Burch, K.~S.}, \bibinfo{author}{Han, J.~H.} \&
  \bibinfo{author}{Ran, Y.}
\newblock \bibinfo{title}{{When Chiral Photons Meet Chiral Fermions:
  Photoinduced Anomalous Hall Effects in Weyl Semimetals}}.
\newblock \emph{\bibinfo{journal}{Physical Review Letters}}
  \textbf{\bibinfo{volume}{116}}, \bibinfo{pages}{026805}
  (\bibinfo{year}{2016}).

\bibitem{Yan:2016eea}
\bibinfo{author}{Yan, Z.} \& \bibinfo{author}{Wang, Z.}
\newblock \bibinfo{title}{{Tunable Weyl Points in Periodically Driven Nodal
  Line Semimetals}}.
\newblock \emph{\bibinfo{journal}{Physical Review Letters}}
  \textbf{\bibinfo{volume}{117}}, \bibinfo{pages}{087402}
  (\bibinfo{year}{2016}).

\bibitem{Yan:2017bv}
\bibinfo{author}{Yan, Z.} \& \bibinfo{author}{Wang, Z.}
\newblock \bibinfo{title}{{Floquet multi-Weyl points in crossing-nodal-line
  semimetals}}.
\newblock \emph{\bibinfo{journal}{Physical Review B}}
  \textbf{\bibinfo{volume}{96}}, \bibinfo{pages}{041206}
  (\bibinfo{year}{2017}).

\bibitem{Zhang:2016di}
\bibinfo{author}{Zhang, X.-X.}, \bibinfo{author}{Ong, T.~T.} \&
  \bibinfo{author}{Nagaosa, N.}
\newblock \bibinfo{title}{{Theory of photoinduced Floquet Weyl semimetal
  phases}}.
\newblock \emph{\bibinfo{journal}{Physical Review B}}
  \textbf{\bibinfo{volume}{94}}, \bibinfo{pages}{235137}
  (\bibinfo{year}{2016}).

\bibitem{Taguchi:2016ef}
\bibinfo{author}{Taguchi, K.}, \bibinfo{author}{Imaeda, T.},
  \bibinfo{author}{Sato, M.} \& \bibinfo{author}{Tanaka, Y.}
\newblock \bibinfo{title}{{Photovoltaic chiral magnetic effect in Weyl
  semimetals}}.
\newblock \emph{\bibinfo{journal}{Physical Review B}}
  \textbf{\bibinfo{volume}{93}}, \bibinfo{pages}{201202}
  (\bibinfo{year}{2016}).

\bibitem{Yao:2017fk}
\bibinfo{author}{Yao, S.}, \bibinfo{author}{Yan, Z.} \& \bibinfo{author}{Wang,
  Z.}
\newblock \bibinfo{title}{{Topological invariants of Floquet systems: General
  formulation, special properties, and Floquet topological defects}}.
\newblock \emph{\bibinfo{journal}{Physical Review B}}
  \textbf{\bibinfo{volume}{96}}, \bibinfo{pages}{195303}
  (\bibinfo{year}{2017}).

\bibitem{Owerre:2017fj}
\bibinfo{author}{Owerre, S.~A.}
\newblock \bibinfo{title}{{Floquet topological magnons}}.
\newblock \emph{\bibinfo{journal}{Journal of Physics Communications}}
  \textbf{\bibinfo{volume}{1}}, \bibinfo{pages}{021002} (\bibinfo{year}{2017}).

\bibitem{Owerre:2018eo}
\bibinfo{author}{Owerre, S.~A.}
\newblock \bibinfo{title}{{Floquet Weyl Magnons in Three-Dimensional Quantum
  Magnets}}.
\newblock \emph{\bibinfo{journal}{Scientific Reports}}
  \textbf{\bibinfo{volume}{8}}, \bibinfo{pages}{10098} (\bibinfo{year}{2018}).

\bibitem{Owerre:2018dz}
\bibinfo{author}{Owerre, S.~A.}
\newblock \bibinfo{title}{{Photoinduced Topological Phase Transitions in
  Topological Magnon Insulators}}.
\newblock \emph{\bibinfo{journal}{Scientific Reports}}
  \textbf{\bibinfo{volume}{8}}, \bibinfo{pages}{4431} (\bibinfo{year}{2018}).

\bibitem{lindner_floquet_2011}
\bibinfo{author}{Lindner, N.~H.}, \bibinfo{author}{Refael, G.} \&
  \bibinfo{author}{Galitski, V.}
\newblock \bibinfo{title}{{Floquet topological insulator in semiconductor
  quantum wells}}.
\newblock \emph{\bibinfo{journal}{Nature Physics}}
  \textbf{\bibinfo{volume}{7}}, \bibinfo{pages}{490--495}
  (\bibinfo{year}{2011}).

\bibitem{Katan:2013hl}
\bibinfo{author}{Katan, Y.~T.} \& \bibinfo{author}{Podolsky, D.}
\newblock \bibinfo{title}{{Modulated Floquet Topological Insulators}}.
\newblock \emph{\bibinfo{journal}{Physical Review Letters}}
  \textbf{\bibinfo{volume}{110}}, \bibinfo{pages}{016802}
  (\bibinfo{year}{2013}).

\bibitem{sie_valley-selective_2014}
\bibinfo{author}{Sie, E.~J.} \emph{et~al.}
\newblock \bibinfo{title}{{Valley-selective optical Stark effect in monolayer
  WS2}}.
\newblock \emph{\bibinfo{journal}{Nature Materials}}  (\bibinfo{year}{2015}).

\bibitem{Claassen:2016ge}
\bibinfo{author}{Claassen, M.}, \bibinfo{author}{Jia, C.},
  \bibinfo{author}{Moritz, B.} \& \bibinfo{author}{Devereaux, T.~P.}
\newblock \bibinfo{title}{{All-optical materials design of chiral edge modes in
  transition-metal dichalcogenides}}.
\newblock \emph{\bibinfo{journal}{Nature Communications}}
  \textbf{\bibinfo{volume}{7}}, \bibinfo{pages}{13074} (\bibinfo{year}{2016}).

\bibitem{Liu:2018dk}
\bibinfo{author}{Liu, H.}, \bibinfo{author}{Sun, J.-T.},
  \bibinfo{author}{Cheng, C.}, \bibinfo{author}{Liu, F.} \&
  \bibinfo{author}{Meng, S.}
\newblock \bibinfo{title}{{Photoinduced Nonequilibrium Topological States in
  Strained Black Phosphorus}}.
\newblock \emph{\bibinfo{journal}{Physical Review Letters}}
  \textbf{\bibinfo{volume}{120}}, \bibinfo{pages}{237403}
  (\bibinfo{year}{2018}).

\bibitem{Wang:2013fe}
\bibinfo{author}{Wang, Y.~H.}, \bibinfo{author}{Steinberg, H.},
  \bibinfo{author}{Jarillo-Herrero, P.} \& \bibinfo{author}{Gedik, N.}
\newblock \bibinfo{title}{{Observation of Floquet-Bloch States on the Surface
  of a Topological Insulator}}.
\newblock \emph{\bibinfo{journal}{Science}} \textbf{\bibinfo{volume}{342}},
  \bibinfo{pages}{453--457} (\bibinfo{year}{2013}).

\bibitem{Mahmood:2016bu}
\bibinfo{author}{Mahmood, F.} \emph{et~al.}
\newblock \bibinfo{title}{{Selective scattering between Floquet-Bloch and
  Volkov states in a topological insulator}}.
\newblock \emph{\bibinfo{journal}{Nature Physics}}
  \textbf{\bibinfo{volume}{12}}, \bibinfo{pages}{306--310}
  (\bibinfo{year}{2016}).

\bibitem{deGiovannini:2016cb}
\bibinfo{author}{de~Giovannini, U.}, \bibinfo{author}{H{\"u}bener, H.} \&
  \bibinfo{author}{Rubio, A.}
\newblock \bibinfo{title}{{Monitoring Electron-Photon Dressing in WSe 2}}.
\newblock \emph{\bibinfo{journal}{Nano Letters}} \textbf{\bibinfo{volume}{16}},
  \bibinfo{pages}{7993--7998} (\bibinfo{year}{2016}).

\bibitem{Thouless:1982kq}
\bibinfo{author}{Thouless, D.~J.}, \bibinfo{author}{Kohmoto, M.},
  \bibinfo{author}{Nightingale, M.~P.} \& \bibinfo{author}{den Nijs, M.}
\newblock \bibinfo{title}{{Quantized Hall Conductance in a Two-Dimensional
  Periodic Potential}}.
\newblock \emph{\bibinfo{journal}{Physical Review Letters}}
  \textbf{\bibinfo{volume}{49}}, \bibinfo{pages}{405--408}
  (\bibinfo{year}{1982}).

\bibitem{2018McIver}
\bibinfo{author}{McIver, J.~W.} \emph{et~al.}
\newblock \bibinfo{title}{Light-induced anomalous hall effect in graphene}.
\newblock \emph{\bibinfo{journal}{arXiv:1811.03522}}  (\bibinfo{year}{2018}).

\bibitem{Sato2019a}
\bibinfo{author}{Sato, S.~A.} \emph{et~al.}
\newblock \bibinfo{title}{Microscopic theory for the light-induced anomalous
  hall effect in graphene}.
\newblock \emph{\bibinfo{journal}{arXiv:XXXX.XXXX}}  (\bibinfo{year}{2019}).

\bibitem{Dehghani:2014jm}
\bibinfo{author}{Dehghani, H.}, \bibinfo{author}{Oka, T.} \&
  \bibinfo{author}{Mitra, A.}
\newblock \bibinfo{title}{{Dissipative Floquet topological systems}}.
\newblock \emph{\bibinfo{journal}{Physical Review B}}
  \textbf{\bibinfo{volume}{90}}, \bibinfo{pages}{195429}
  (\bibinfo{year}{2014}).

\bibitem{Dehghani:2015gz}
\bibinfo{author}{Dehghani, H.}, \bibinfo{author}{Oka, T.} \&
  \bibinfo{author}{Mitra, A.}
\newblock \bibinfo{title}{{Out-of-equilibrium electrons and the Hall
  conductance of a Floquet topological insulator}}.
\newblock \emph{\bibinfo{journal}{Physical Review B}}
  \textbf{\bibinfo{volume}{91}}, \bibinfo{pages}{155422}
  (\bibinfo{year}{2015}).

\bibitem{Sato2019b}
\bibinfo{author}{Sato, S.~A.} \emph{et~al.}
\newblock \bibinfo{title}{Light-induced anomalous hall effect in massless dirac
  fermion systems and topological insulators with dissipation}.
\newblock \emph{\bibinfo{journal}{arXiv:XXXX.XXXX}}  (\bibinfo{year}{2019}).

\bibitem{Murakami:2017ht}
\bibinfo{author}{Murakami, Y.}, \bibinfo{author}{Tsuji, N.},
  \bibinfo{author}{Eckstein, M.} \& \bibinfo{author}{Werner, P.}
\newblock \bibinfo{title}{{Nonequilibrium steady states and transient dynamics
  of conventional superconductors under phonon driving}}.
\newblock \emph{\bibinfo{journal}{Physical Review B}}
  \textbf{\bibinfo{volume}{96}}, \bibinfo{pages}{045125}
  (\bibinfo{year}{2017}).

\bibitem{Hubener:2018id}
\bibinfo{author}{H{\"u}bener, H.}, \bibinfo{author}{de~Giovannini, U.} \&
  \bibinfo{author}{Rubio, A.}
\newblock \bibinfo{title}{{Phonon Driven Floquet Matter}}.
\newblock \emph{\bibinfo{journal}{Nano Letters}} \textbf{\bibinfo{volume}{18}},
  \bibinfo{pages}{1535--1542} (\bibinfo{year}{2018}).

\bibitem{Garate:2013er}
\bibinfo{author}{Garate, I.}
\newblock \bibinfo{title}{{Phonon-Induced Topological Transitions and
  Crossovers in Dirac Materials}}.
\newblock \emph{\bibinfo{journal}{Physical Review Letters}}
  \textbf{\bibinfo{volume}{110}}, \bibinfo{pages}{046402}
  (\bibinfo{year}{2013}).

\bibitem{Wang:2017by}
\bibinfo{author}{Wang, L.-L.} \emph{et~al.}
\newblock \bibinfo{title}{{Phonon-induced topological transition to a type-II
  Weyl semimetal}}.
\newblock \emph{\bibinfo{journal}{Physical Review B}}
  \textbf{\bibinfo{volume}{95}}, \bibinfo{pages}{165114}
  (\bibinfo{year}{2017}).

\bibitem{Antonius:2016gj}
\bibinfo{author}{Antonius, G.} \& \bibinfo{author}{Louie, S.~G.}
\newblock \bibinfo{title}{{Temperature-Induced Topological Phase Transitions:
  Promoted versus Suppressed Nontrivial Topology}}.
\newblock \emph{\bibinfo{journal}{Physical Review Letters}}
  \textbf{\bibinfo{volume}{117}}, \bibinfo{pages}{246401}
  (\bibinfo{year}{2016}).

\bibitem{Lanzara:2001kp}
\bibinfo{author}{Lanzara, A.} \emph{et~al.}
\newblock \bibinfo{title}{{Evidence for ubiquitous strong
  electron{\textendash}phonon coupling in high-temperature superconductors}}.
\newblock \emph{\bibinfo{journal}{Nature}} \textbf{\bibinfo{volume}{412}},
  \bibinfo{pages}{510--514} (\bibinfo{year}{2001}).

\bibitem{Lee:2014bw}
\bibinfo{author}{Lee, J.~J.} \emph{et~al.}
\newblock \bibinfo{title}{{Interfacial mode coupling as the origin of the
  enhancement of <i>T</i><sub>c</sub> in FeSe films on SrTiO<sub>3 </sub>}}.
\newblock \emph{\bibinfo{journal}{Nature}} \textbf{\bibinfo{volume}{515}},
  \bibinfo{pages}{nature13894--248} (\bibinfo{year}{2014}).

\bibitem{Antonius:2015fj}
\bibinfo{author}{Antonius, G.} \emph{et~al.}
\newblock \bibinfo{title}{{Dynamical and anharmonic effects on the
  electron-phonon coupling and the zero-point renormalization of the electronic
  structure}}.
\newblock \emph{\bibinfo{journal}{Physical Review B}}
  \textbf{\bibinfo{volume}{92}}, \bibinfo{pages}{085137}
  (\bibinfo{year}{2015}).

\bibitem{Story:2014hn}
\bibinfo{author}{Story, S.~M.}, \bibinfo{author}{Kas, J.~J.},
  \bibinfo{author}{Vila, F.~D.}, \bibinfo{author}{Verstraete, M.~J.} \&
  \bibinfo{author}{Rehr, J.~J.}
\newblock \bibinfo{title}{{Cumulant expansion for phonon contributions to the
  electron spectral function}}.
\newblock \emph{\bibinfo{journal}{Physical Review B}}
  \textbf{\bibinfo{volume}{90}}, \bibinfo{pages}{195135}
  (\bibinfo{year}{2014}).

\bibitem{Moser:2013bf}
\bibinfo{author}{Moser, S.} \emph{et~al.}
\newblock \bibinfo{title}{{Tunable Polaronic Conduction in Anatase <span
  class="aps-inline-formula"><math xmlns="http://www.w3.org/1998/Math/MathML"
  display="inline"><msub><mi>TiO</mi><mn>2</mn></msub></math></span>}}.
\newblock \emph{\bibinfo{journal}{Physical Review Letters}}
  \textbf{\bibinfo{volume}{110}}, \bibinfo{pages}{196403}
  (\bibinfo{year}{2013}).

\bibitem{Chen:2015fz}
\bibinfo{author}{Chen, C.}, \bibinfo{author}{Avila, J.},
  \bibinfo{author}{Frantzeskakis, E.}, \bibinfo{author}{Levy, A.} \&
  \bibinfo{author}{Asensio, M.~C.}
\newblock \bibinfo{title}{{Observation of a two-dimensional liquid of
  Fr{\"o}hlich polarons at the bare SrTiO<sub>3</sub> surface}}.
\newblock \emph{\bibinfo{journal}{Nature Communications}}
  \textbf{\bibinfo{volume}{6}}, \bibinfo{pages}{8585} (\bibinfo{year}{2015}).

\bibitem{Cancellieri:2016fw}
\bibinfo{author}{Cancellieri, C.} \emph{et~al.}
\newblock \bibinfo{title}{{Polaronic metal state at the
  LaAlO<sub>3</sub>/SrTiO<sub>3</sub> interface}}.
\newblock \emph{\bibinfo{journal}{Nature Communications}}
  \textbf{\bibinfo{volume}{7}}, \bibinfo{pages}{10386} (\bibinfo{year}{2016}).

\bibitem{Wang:2016de}
\bibinfo{author}{Wang, Z.} \emph{et~al.}
\newblock \bibinfo{title}{{Tailoring the nature and strength of
  electron{\textendash}phonon interactions in the SrTiO<sub><b>3</b></sub>(001)
  2D electron~liquid}}.
\newblock \emph{\bibinfo{journal}{Nature Materials}}
  \textbf{\bibinfo{volume}{15}}, \bibinfo{pages}{835--839}
  (\bibinfo{year}{2016}).

\bibitem{Verdi:2017bx}
\bibinfo{author}{Verdi, C.}, \bibinfo{author}{Caruso, F.} \&
  \bibinfo{author}{Giustino, F.}
\newblock \bibinfo{title}{{Origin of the crossover from polarons to Fermi
  liquids in transition metal oxides}}.
\newblock \emph{\bibinfo{journal}{Nature Communications}}
  \textbf{\bibinfo{volume}{8}}, \bibinfo{pages}{15769} (\bibinfo{year}{2017}).

\bibitem{Gerber:2017bm}
\bibinfo{author}{Gerber, S.} \emph{et~al.}
\newblock \bibinfo{title}{{Femtosecond electron-phonon lock-in by photoemission
  and x-ray free-electron laser}}.
\newblock \emph{\bibinfo{journal}{Science}} \textbf{\bibinfo{volume}{357}},
  \bibinfo{pages}{71--75} (\bibinfo{year}{2017}).

\bibitem{Shin:2018ek}
\bibinfo{author}{Shin, D.} \emph{et~al.}
\newblock \bibinfo{title}{{Phonon-driven spin-Floquet magneto-valleytronics in
  MoS 2}}.
\newblock \emph{\bibinfo{journal}{Nature Communications}}
  \textbf{\bibinfo{volume}{9}}, \bibinfo{pages}{638} (\bibinfo{year}{2018}).

\bibitem{Guzzo:2011dc}
\bibinfo{author}{Guzzo, M.} \emph{et~al.}
\newblock \bibinfo{title}{{Valence Electron Photoemission Spectrum of
  Semiconductors: <i>Ab~Initio</i> Description of Multiple Satellites}}.
\newblock \emph{\bibinfo{journal}{Physical Review Letters}}
  \textbf{\bibinfo{volume}{107}}, \bibinfo{pages}{166401}
  (\bibinfo{year}{2011}).

\bibitem{Guzzo:2014gu}
\bibinfo{author}{Guzzo, M.} \emph{et~al.}
\newblock \bibinfo{title}{{Multiple satellites in materials with complex
  plasmon spectra: From graphite to graphene}}.
\newblock \emph{\bibinfo{journal}{Physical Review B}}
  \textbf{\bibinfo{volume}{89}}, \bibinfo{pages}{085425}
  (\bibinfo{year}{2014}).

\bibitem{Kas:2014dj}
\bibinfo{author}{Kas, J.~J.}, \bibinfo{author}{Rehr, J.~J.} \&
  \bibinfo{author}{Reining, L.}
\newblock \bibinfo{title}{{Cumulant expansion of the retarded one-electron
  Green function}}.
\newblock \emph{\bibinfo{journal}{Physical Review B}}
  \textbf{\bibinfo{volume}{90}}, \bibinfo{pages}{085112}
  (\bibinfo{year}{2014}).

\bibitem{Offi:2007hq}
\bibinfo{author}{Offi, F.} \emph{et~al.}
\newblock \bibinfo{title}{{Comparison of hard and soft x-ray photoelectron
  spectra of silicon}}.
\newblock \emph{\bibinfo{journal}{Physical Review B}}
  \textbf{\bibinfo{volume}{76}}, \bibinfo{pages}{085422}
  (\bibinfo{year}{2007}).

\bibitem{Caruso:2015fj}
\bibinfo{author}{Caruso, F.}, \bibinfo{author}{Lambert, H.} \&
  \bibinfo{author}{Giustino, F.}
\newblock \bibinfo{title}{{Band Structures of Plasmonic Polarons}}.
\newblock \emph{\bibinfo{journal}{Physical Review Letters}}
  \textbf{\bibinfo{volume}{114}}, \bibinfo{pages}{146404}
  (\bibinfo{year}{2015}).

\bibitem{Lischner:2015fd}
\bibinfo{author}{Lischner, J.} \emph{et~al.}
\newblock \bibinfo{title}{{Satellite band structure in silicon caused by
  electron-plasmon coupling}}.
\newblock \emph{\bibinfo{journal}{Physical Review B}}
  \textbf{\bibinfo{volume}{91}}, \bibinfo{pages}{205113}
  (\bibinfo{year}{2015}).

\end{thebibliography}

\end{document}